\documentclass[lettersize,journal]{IEEEtran}
\IEEEoverridecommandlockouts
\usepackage{cite}
\usepackage{amsmath,amssymb,amsfonts}
\usepackage{algorithmic}
\usepackage{subcaption}
\usepackage{glossaries}
\usepackage{mathtools}
\usepackage{algorithm}
\usepackage{makecell}
\usepackage{graphicx}
\usepackage{textcomp}
\usepackage{comment}
\usepackage{xcolor}
\usepackage{soul}
\usepackage{svg}
\usepackage{bm}
\def\BibTeX{{\rm B\kern-.05em{\sc i\kern-.025em b}\kern-.08em
    T\kern-.1667em\lower.7ex\hbox{E}\kern-.125emX}}

\allowdisplaybreaks

\newacronym{DAG}{DAG}{Directed Acyclic Graph}
\newacronym{FCFS}{FCFS}{First-Come First-Served}
\newacronym{SFCs}{SFCs}{Service Function Chains}
\newacronym{SLA}{SLA}{Service-Level Agreement}
\newacronym{AWS}{AWS}{Amazon Web Services}

\newcommand{\E}{\mathbb{E}}
\newcommand{\Edges}{\mathcal{E}}
\newcommand{\R}{\mathcal{R}}
\newcommand{\K}{\mathcal{K}}
\newcommand{\X}{\mathcal{X}}
\newcommand{\V}{\mathcal{V}}
\newcommand{\G}{\mathcal{G}}
\newcommand{\eps}{\varepsilon}
\newcommand{\GR}{\textnormal{GR} }
\newcommand{\SR}{\textnormal{SR} }

\newcommand{\Pone}{$\mathcal{P}_1$}
\newcommand{\Ptwo}{$\mathcal{P}_2$}
    
\begin{document}

\title{
SPARQ: An Optimization Framework for the Distribution of AI-Intensive Applications under Non-Linear Delay Constraints\\
\thanks{This work was partially supported by the European Union under the Italian National Recovery and Resilience Plan (NRRP) of NextGenerationEU, partnership on “Telecommunications of the Future” (PE00000001 - program “RESTART”).

P. Spadaccino, P. Di Lorenzo, and S. Barbarossa are with the Department of Information Engineering, Electronics, and Telecommunications of Sapienza University, 00184 Roma, Italy (e-mail: \{name.surname\}@uniroma1.it).

A. M. Tulino is with the Department of Electrical Engineering, Università degli Studi di Napoli Federico II, 80138 Napoli, Italy, and with the Electrical and Computer Engineering Department, New York University, USA (e-mail: antoniamaria.tulino@unina.it),

J. Llorca is with the Department of Information Science and Engineering, Università degli Studi di Trento, Italy, and with the Electrical and Computer Engineering Department, New York University, USA (e-mail: jaime.llorca@unitn.it)
}
}

\author{Pietro Spadaccino, Paolo Di Lorenzo, Sergio Barbarossa, Antonia M. Tulino, Jaime Llorca}%

\maketitle

\begin{abstract}
Next-generation real-time compute-intensive applications, such as extended reality, multi-user gaming, and autonomous transportation, are increasingly composed of heterogeneous AI-intensive functions with diverse resource requirements and stringent latency constraints. While recent advances have enabled very efficient algorithms for joint service placement, routing, and resource allocation for increasingly complex applications, current models fail to capture the non-linear relationship between delay and resource usage that becomes especially relevant in AI-intensive workloads. In this paper, we extend the {\em cloud network flow} optimization framework to support queuing-delay-aware orchestration of distributed AI applications over edge-cloud infrastructures. We introduce two execution models—Guaranteed-Resource (GR) and Shared-Resource (SR)—that more accurately capture how computation and communication delays emerge from system-level resource constraints. These models incorporate M/M/1 and M/G/1 queue dynamics to represent dedicated and shared resource usage, respectively. The resulting optimization problem is non-convex due to the non-linear delay terms. To overcome this, we develop SPARQ, an iterative approximation algorithm that decomposes the problem into two convex sub-problems, enabling joint optimization of service placement, routing, and resource allocation under nonlinear delay constraints. 
Simulation results demonstrate that the SPARQ not only offers a more faithful representation of system delays, but also substantially improves resource efficiency and the overall cost-delay tradeoff compared to existing state-of-the-art methods.
\end{abstract}


\section{Introduction}
\label{sec:intro}
The proliferation of distributed cloud architectures, such as Multi-access Edge Computing (MEC) and Fog Computing, has enabled innovative service models capable of addressing the stringent latency and computational demands of AI and next-generation applications \cite{ex1}, \cite{ex2}. These applications, including augmented reality, autonomous driving, and industrial automation, require the efficient coordination of communication, computation, and storage resources. \gls{SFCs} or, more generally, service graphs, which define sequences of interdependent service functions, are crucial for meeting these demands. Properly embedding these SFCs across distributed cloud infrastructures is vital for optimizing performance while minimizing resource consumption. Existing research has largely focused on static and dynamic placement and optimization of SFCs, treating the orchestration of service chains as a combinatorial optimization problem aimed at minimizing costs or maximizing throughput.
Due to the complexity and non-linearity of the system, however, approximations have been used which, depending on the case, can be too rough or stringent.
Moreover, SFC optimization must consider the growing data requirements of these applications, along with the inherent non-linearity of network delays under stochastic traffic conditions.

\begin{figure}
    \centering
    \includegraphics[width=0.9\linewidth]{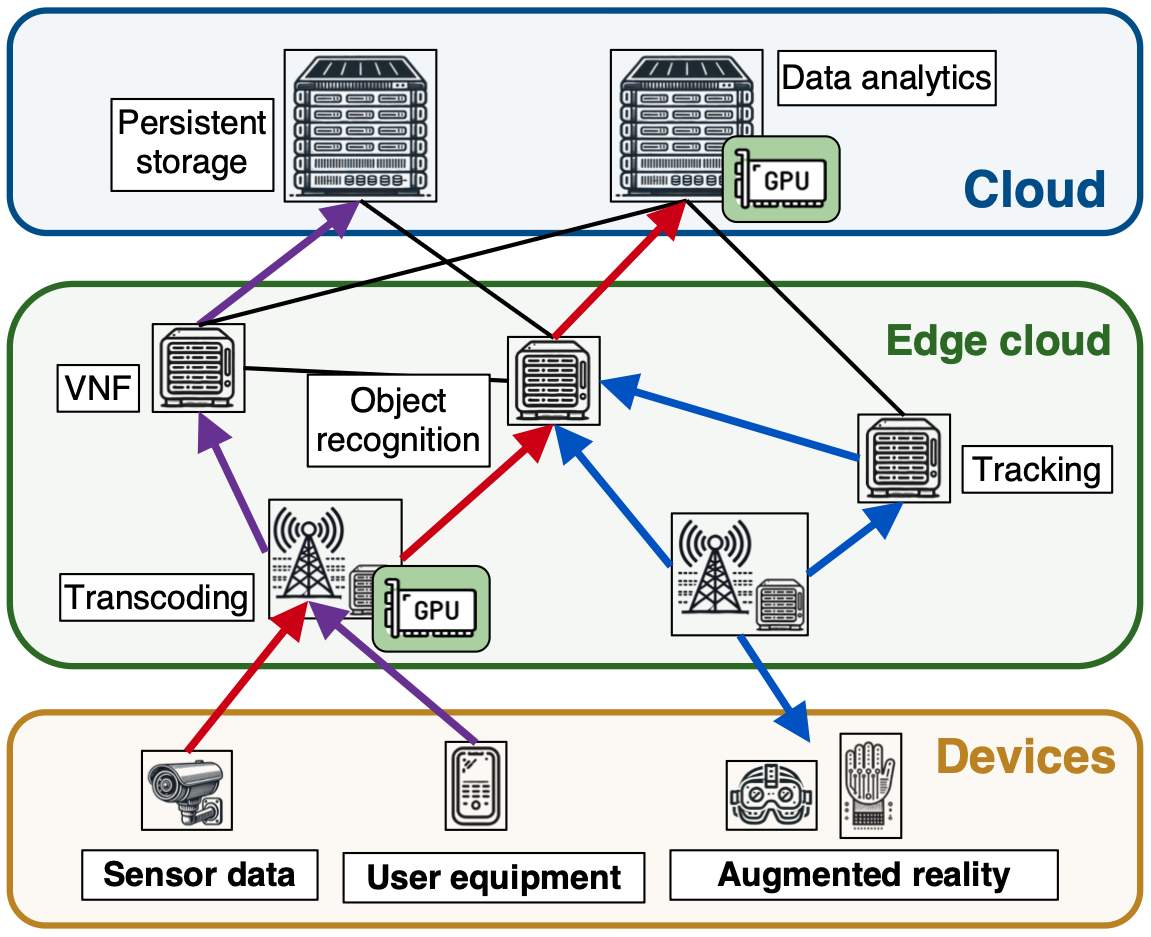}
    \caption{Edge-cloud architecture. Various service chains running over a shared distributed computing network. 
    Network elements in the edge-cloud architecture may be equipped with CPU and/or GPU processing capabilities, to accommodate AI workloads.}
    \label{fig:edge-cloud-flow}
\end{figure}

To address the complexities and requirements of \gls{SFCs}, edge-cloud architectures are introduced. Fig. \ref{fig:edge-cloud-flow} illustrates an example of this architecture, where services from multiple \gls{SFCs} are distributed across various nodes in the network.
In our modeling framework, an SFC is represented by a \gls{DAG}, whose vertices represent service functions or tasks, and edges represent  \textit{commodities} or 
data streams flowing through the network.

Numerous applications can benefit from such an architecture, including tasks like image and text recognition, data analytics, and video stream processing. Decisions around function placement, flow routing, and resource allocation 
must carefully consider the cost of operating the required communication and computations resources, as well as the delay incurred in transmitting data along communication links, and in processing data by the service functions hosted at the edge-cloud computation resources. 

\subsection{Motivation}

When designing the placement, routing, and resource allocation of distributed applications, it is also crucial to consider the underlying computing architecture and deployment constraints. This is particularly relevant for AI-based applications, which impose diverse and stringent computational requirements.
AI workloads typically involve a combination of CPU-intensive tasks—such as data pre-processing, feature extraction, and general computations—and GPU-intensive tasks, which primarily involve executing deep learning models for inference or training. The latter often require substantial VRAM capacity, as modern AI models tend to be large \cite{deepseek}, fully occupying the available GPU memory for the duration of execution.
When an AI model is loaded onto a GPU-equipped node, it fully occupies the GPU memory, preventing concurrent execution of other models on the same device until processing is complete. Conversely, CPU-only tasks may either serve as AI pre-processing steps or represent general-purpose computations unrelated to AI, such as database queries, logging, or application control logic.
Such operations typically fit comfortably within the system RAM, enabling multiple processes to reside simultaneously in memory. The operating system efficiently manages these tasks by rapidly performing context-switching, which involves temporarily pausing one task and resuming another. Because these context switches occur at very high frequencies, the user perceives that multiple CPU tasks are executed concurrently, despite the CPU physically processing instructions sequentially.

\begin{figure}[t]
    \centering
    \begin{subfigure}[b]{0.98\linewidth}
        \centering
        \includegraphics[width=\linewidth]{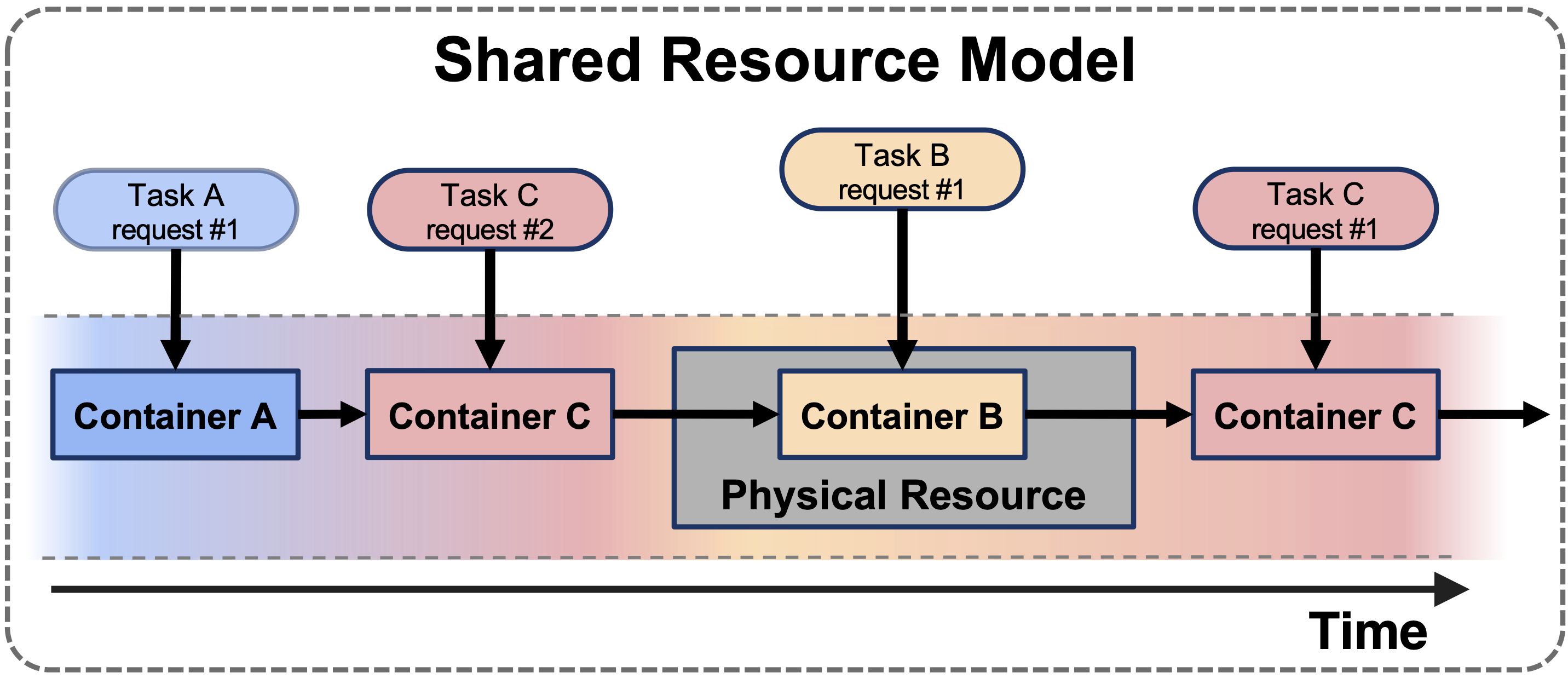}
        \caption{Heterogeneous tasks access the same physical resource (e.g., a GPU) without resource-level isolation. Different tasks 
        compete for the shared resource, leading to interdependent delays.}
        \label{fig:shared_res}
        \vspace{0.4cm}
    \end{subfigure}
    \begin{subfigure}[b]{0.98\linewidth}
        \centering
        \includegraphics[width=\linewidth]{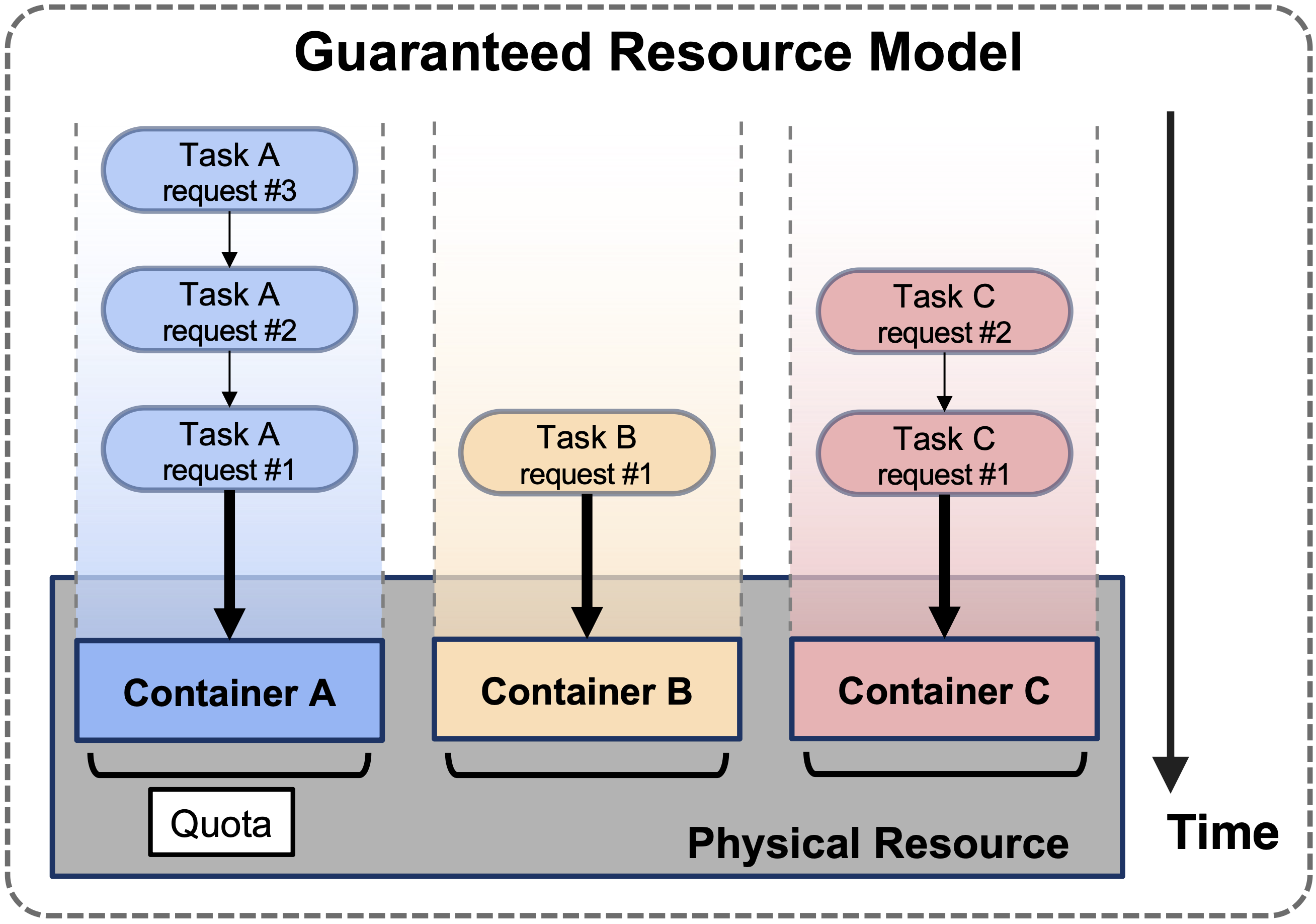}
        \caption{Heterogeneous tasks are allocated a fixed quota of the physical resource (e.g., CPU time slice or reserved GPU memory), enabling isolated execution. Tasks are processed independently within their assigned quota, leading to non-interdependent delays.}
        \label{fig:guaranteed_res}
    \end{subfigure}
    \caption{Model of a shared resource over time (a) and a resource with guaranteed quotas over time (b).}
    \label{fig:shared_guaranteed_res}
\end{figure}

Expanding the discussion beyond just CPU and GPU, we can identify two general classes of computational workloads, schematized in Fig. \ref{fig:shared_guaranteed_res}. The first class, as reported in Fig. \ref{fig:shared_res}, comprises tasks that fully utilize the available computational resources, effectively blocking or monopolizing them during execution. Such tasks prevent simultaneous use of these resources by other tasks, leading to a scenario where concurrent execution or interleaving of multiple tasks is not possible. In this case, the same physical resource is \textit{shared} across several tasks. Examples include tasks running large AI models fully occupying a GPU's VRAM, or specialized computations on dedicated hardware like FPGA accelerators. The second class includes tasks that only partially utilize the available resources, as reported in Fig. \ref{fig:guaranteed_res}, allowing multiple such tasks to coexist simultaneously without resource contention. Due to their limited resource requirements, these tasks can efficiently run alongside others within the same computational environment, benefiting from the operating system's task scheduling and multiplexing capabilities. Examples include lightweight CPU-bound processes, lightweight GPU processing, containerized microservices, or any task executed within \textit{guaranteed} resource quotas that ensure minimal interference with others.

\subsection{Related work}
The optimization of \gls{SFCs} in distributed cloud networks has been extensively studied in recent years, particularly in the context of minimizing  resource operational cost while meeting quality of service (QoS) requirements (e.g., latency). 
Several models have been proposed to capture the flow of requests and the placement of services within distributed computing networks. Among these, the Cloud Network Flow (CNF) model introduced in \cite{barcelo2016comp,marcelo,poularakis2020} stands out. This model represents Service Function Chains (SFCs) as Directed Acyclic Graphs (DAGs), allowing for an efficient representation of complex service workflows. Distributed computing networks are then modeled via {\em cloud-augmented graphs}, where in addition to communication links representing data transmission capabilities, there are production, consumption, and computation links representing the capability to produce, consume, and process data throughout he network. By leveraging this structure, the model enables the joint optimization of placement, routing, and resource allocation by solving a single flow problem on the cloud-augmented graph.
The CNF model has been widely adopted for the study of distributed cloud-edge networks with embedded service DAGs, accommodating both static optimization and dynamic control scenarios.
Static service optimization, including placement, routing, and resource allocation, has been explored in works such as \cite{barcelo2016comp, marcelo, poularakis2020,Poularakis2020Mobihoc, mauro,Llorca2024CNF}, addressing varying levels of service complexity and resource demands. However, these studies do not account for queueing delays.
In contrast, studies like \cite{fengDCNC, zhang2021} focus on dynamic service control policies that optimize routing, processing, and scheduling decisions based on local real-time observations. 
While these approaches integrate queueing systems and account for associated delays, they primarily aim to achieve bounded average delay through queue stability, without explicitly guaranteeing end-to-end service latency.

Various studies have proposed models to address the challenge of mapping service chains onto network resources, often leveraging queuing theory to capture traffic dynamics. For instance, \cite{pezone2022} models communication links as queues and optimizes queue throughput by stabilizing the Lyapunov drift function. While this approach effectively addresses resource allocation, it does not consider the placement or routing of SFCs.
In \cite{cai2023joint}, the authors explore dynamic service control in edge networks, addressing both live and static data streams. They employ a graph-based model to represent network processing and transmission resources, with queues managing Poisson-distributed arrival rates. This approach facilitates the handling of dynamic traffic flows and ensures throughput-optimal performance.
In \cite{9500780}, the authors develop a fully distributed packet processing and routing policy for multicast services. Traffic across multiple network links is modeled using queues, enabling dynamic control of packet processing and duplication. This method captures the behavior of data flows traversing the network, ensuring throughput-optimal routing and efficient resource utilization. Also, \cite{9040668} introduces a dynamic computation offloading strategy tailored for MEC environments, with a focus on ultra-reliable low-latency communications (URLLC). The approach utilizes queuing models to dynamically manage communication and computation tasks, optimizing the joint allocation of radio and computational resources. By ensuring that the sum of communication and computation queue lengths remains within a probabilistic bound, their algorithm balances service delay and energy consumption while adhering to reliability constraints. This stochastic optimization approach effectively addresses latency requirements under dynamic network conditions.

In addition to focusing solely on modeling and embeddings, several works in the literature also take into account practical and implementation-related challenges. These studies address, among other things, the issue of memory limitations in network routers, specifically Ternary Content Addressable Memories (TCAMs) \cite{tcam1, tcam2}, which are crucial for performing SFC placement and request classification in NFV/SDN environments. Authors in \cite{polverini2021}, by proposing an offloading strategy, use an Integer Linear Programming (ILP) formulation and heuristic approaches to efficiently handle SFC requests under TCAM size constraints. This approach is more practical as it directly addresses the hardware limitations that impact SFC deployment, offering a scalable solution to increase the number of classified requests without significantly affecting network Quality of Service (QoS). Moreover, in \cite{8611305}, the authors address the critical challenge of efficiently placing Virtual Network Functions (VNFs) \gls{SFCs} and allocating resources in 5G networks, specifically tailored to vertical industries. They propose a queuing-based model to map service requirements into decisions about VNF placement, CPU assignment, and traffic routing.
Their strategy is particularly suited for practical implementations in 5G networks, where it dynamically adjusts resource allocation based on traffic loads, ensuring low latency and efficient use of network resources.

\subsection{Main contributions}

In this paper, we propose \textbf{SPARQ} (Service Placement, Resource Allocation and Routing with Queue-Aware Delays), a {\bf comprehensive optimization framework to  jointly optimize service function placement, flow routing, and computation-communication resource allocation} with the goal of minimizing overall operational cost while guaranteeing resource capacity and end-to-end latency constraints.  
Importantly, {\bf latency constraints capture the critical nonlinearities associated with the delay experienced by next-generation AI-intensive applications} consuming heterogeneous shared and guaranteed resources. 
 
In particular, we propose {\bf two resource models, named Guaranteed-Resource (GR) and Shared-Resource (SR) models}, which offer a more accurate representation of how tasks are executed on modern hardware. 
The GR model ensures that tasks are allocated dedicated resources, preventing contention and allowing each task to be processed independently. In contrast, the SR model models scenarios where tasks must share a resource, and the delays experienced by each task depend on the usage of the resource by other tasks. This distinction is crucial to accurately model real-world scenarios, especially in the context of emerging applications with an increasing presence of AI workloads, where resource contention (e.g., GPU memory) plays a significant role in overall system performance.

To incorporate these models into the CNF framework, we propose a queue-based modeling approach for the computation and communication links within the cloud-augmented graph. This approach models {\bf each communication and computation system} (links in the cloud-augmented graph) 
{\bf as a  mixture of M/G/1 and M/M/1 queues} to accurately capture the delay experienced by service commodities (or data streams) flowing through the network. 
Our model precisely expresses delay as a function of service placement, incoming request volume, and the resources available to serve these requests.
Unlike the private delay model\footnote{In the private delay, model, delay is assumed constant with respect to allocated resources, as long as allocated resources are above average load, hence ignoring the queuing and congestion effects due to stochastic traffic variations.} assumed in previous approaches, this flexible queue-based approach more accurately captures service  delays in stochastic AI-intensive applications. 
Moreover, prior works use queuing models primarily for communication links. In our approach, leveraging the cloud-augmented graph, {\bf we extend queue delay modeling to all network components} (communication and computation resources) creating a more comprehensive and realistic representation of the compute-communication system.

However, direct optimization using this model is challenging due to its non-convex formulation. To address this, we develop several {\bf approximation and convexification techniques} specific to our model. These methods are {\bf integrated into a single iterative algorithm that jointly optimizes service placement, routing, and resource allocation with accurate end-to-end latency constraints}.

To assess the performance of the proposed framework, we perform extensive simulations on representative edge-cloud deployment scenarios and realistic traffic patterns. 
Simulation results demonstrate that the approximated delay model closely mirrors the \textit{a-posteriori} measured delay, validating the accuracy of the approximation and confirming its suitability for solving the optimization problem without significant loss in solution fidelity. Moreover, the proposed approach consistently outperforms baseline methods in both cost efficiency and delay compliance, yielding a substantially improved cost-delay tradeoff. Taken together, these results confirm the practical viability of our method and clearly establish its advantage over existing solutions for managing complex, latency-sensitive AI workloads.

\section{System Model}

\begin{figure}[t]
    \centering
    \fbox{\includegraphics[width=0.44\linewidth]{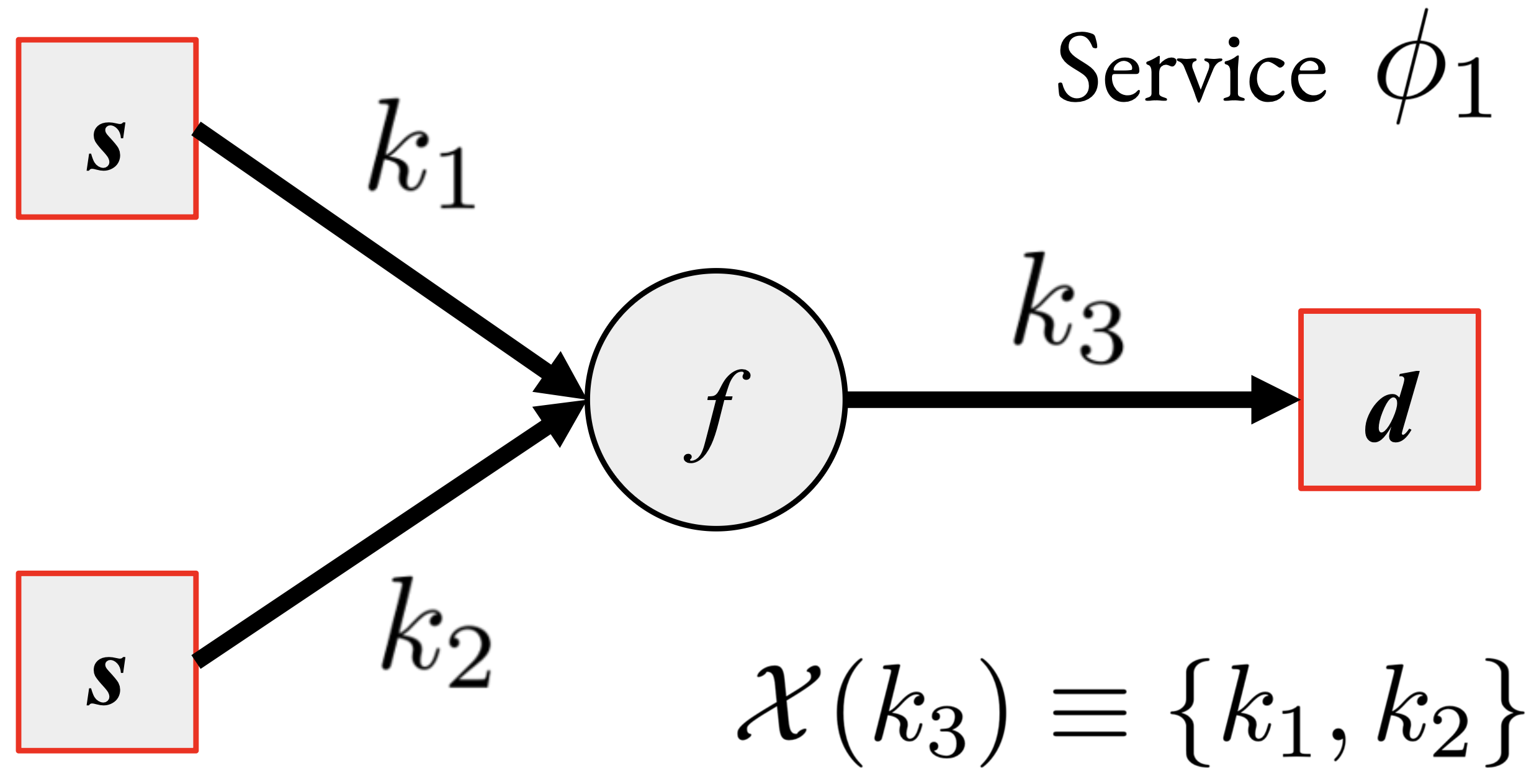}}
    \caption{Example of a service graph. Commodity $k_3$ is produced requiring as input commodities $k_1, k_2$.}
    \label{fig:figurina}
\end{figure}

In this section, we formalize the network and service models that underpin the Cloud Network Flow (CNF) framework for optimizing the deployment of distributed services in edge-cloud environments.

At the core of this framework lies a \textit{multi-commodity-chain flow} problem~\cite{baiocchi2020}, where each commodity~$k$ represents a data flow associated with a specific service component. In the classical multi-commodity-flow formulation, a commodity is defined by a fixed source-destination pair~$(s, d)$, corresponding to a flow of data between two network nodes. 

In the CNF setting, however, this notion is extended to capture the flexibility required by modern service function chains and AI workflows. In particular, some commodities correspond to intermediate service stages whose execution locations are not predefined. For these flows, their source and/or destination nodes are treated as decision variables and are resolved by the optimization process, which jointly determines the placement of service components and the routing of data between them.
In this context, we say that commodity~$k$ is {\em produced} at a node~$u$ either because $u$ is the origin of~$k$ (e.g., a data source such as a sensor), or because $k$ is produced as the result of a computation at~$u$ involving one or more input commodities. 
Similarly, we say $k$ is {\em consumed} at a node~$v$ either because $v$ is the final destination of the data (e.g., a user device or data sink), or because $k$ is used as input to a computation at node $v$ that generates new downstream commodities.

Each commodity represents a distinct data stream (or data flow) and is associated with possibly heterogeneous communication and computation requirements (e.g., bandwidth, latency sensitivity, processing load), thereby reflecting the diverse service-level demands of modern applications.

To effectively capture the interplay between the physical infrastructure, the service logic, and the associated data flows, the CNF framework introduces three distinct graph abstractions, each serving a complementary role in the overall modeling approach: 
i) the \textit{network graph}, which represents the physical infrastructure of the network; ii) the \textit{service graph}, representing the computations required by the application; iii) and the \textit{cloud-augmented} graph, which extends the network graph with additional auxiliary nodes added to represent computation, production and consumption operations.
The network graph is modeled as $\mathcal{G} = (\mathcal{V}, \mathcal{E})$, where $\mathcal{V}$ and $\mathcal{E}$ represent the set of network nodes and links, respectively. For any two nodes $u, v \in \mathcal{V}$, an link $(u, v) \in \mathcal{E}$ indicates a network connection.

As for the service graph, we define $\mathcal{K}$ to be the set of all commodities. A service $\phi \in \Phi$ is modeled as a \gls{DAG}, where edges correspond to the commodities required by the service and vertices represent processing steps, referred to as \textit{service functions}. Specifically, each commodity $k \in \phi$ represents a data flow between two service functions and is therefore associated with a directed edge in the \gls{DAG}. This edge indicates that the output of one function serves as the input to the next. 
Considering commodities $k, j \in \mathcal{K}$, if commodity $j$ is required to produce commodity $k$, this dependency is expressed as $j \in \mathcal{\X}(k)$.
We call $k$ a source commodity if it is generated with no inputs $\X(k)=\emptyset$ in a network node $s(k) \in V$, and denote by $\mathcal K^s\in\mathcal K$ the set of source commodities. Similarly, we call $k$ a destination commodity if it is consumed without generating any output at node $d(k) \in V$, and denote by $\mathcal K^d\in\mathcal K$ the set of destination commodities.

Embedding a service $\phi$ in the network graph means mapping each service function to a physical node ({\em function placement}) and each commodity to a feasible path between them ({\em data routing}). Moreover, we note that multiple services ${\phi_1, \phi_2, \ldots} \in \Phi$ may require to be embedded within the same network graph. We use $\phi(k)$ to denote the service  associated with commodity $k$. An example of a simple service graph is reported in Fig. \ref{fig:figurina}.

\begin{figure}[t]
    \centering
    \includegraphics[width=0.95\linewidth]{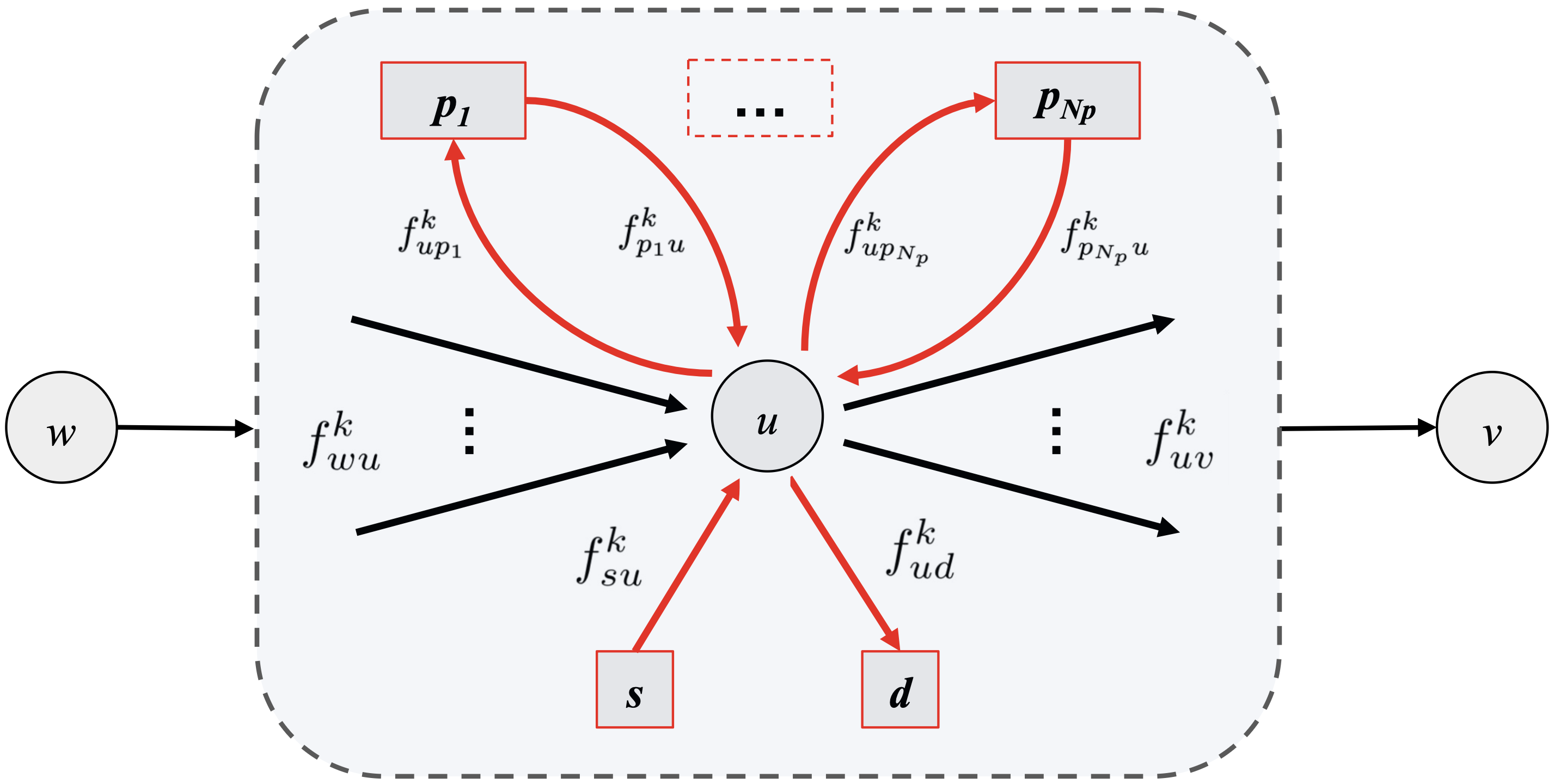}
    \caption{Augmented graph node. For each node $u \in \V$, virtual nodes are created, representing production, consumption, and computation capabilities. The augmentation enables the formulation of the service graph embedding problem within the augmented graph as a single network flow problem.}
    \label{fig:augmented_node}
\end{figure}

To perform an efficient embedding, we make use of the cloud-augmented graph model \cite{barcelo2016comp,marcelo,poularakis2020}. The augmented graph of $\G$ is represented by $\mathcal{G}^a = (\mathcal{V}^a, \mathcal{E}^a)$. It is constructed by creating additional nodes for each communication node $u \in \mathcal{V}$. These nodes are denoted by $p$, $s$, and $d$, modeling the computation, production, and consumption capabilities of node $u$, respectively, as illustrated in Fig. \ref{fig:augmented_node}. 
In general, a node $u\in\mathcal V$ may have more than one computation node: in our model, the computation can be carried out by different types of computational systems $\{p_1, p_2, \dots, p_{N}\}$. These computational systems could, for example, be a specific type of \gls{AWS} EC2 machine (e.g., small or xl), or a machine equipped with additional hardware such as a GPUs or FPGAs. 
A computational system $p_i$ can provide one or more physical resources in $\R=\{r_1, r_2, \dots \}$, representing, for example, CPU, GPU or FPGA capabilities.
Node $u$ can offer zero or more computational systems $\{p_i, p_j, \dots\}$, with computation links $(u,p_i)$ and $(u,p_i) \in \mathcal{E}^a$ existing if node $u$ can offer system $p_i$.


Depending on the roles of nodes \( u \) and \( v \), an edge \( (u,v) \in \mathcal{E}^a \) may represent a communication link, a computation link, or a data transfer to or from storage. This augmentation of the network graph enables the formulation of the service graph embedding problem as a \emph{network flow problem} \cite{barcelo2016comp, cnf2}, making it suitable for tractable optimization through established flow-based techniques.

We define binary {\bf flow variables} \( f^k_{uv} \in \{0,1\} \) to indicate whether commodity \( k \in \mathcal{K} \) is assigned to link \( (u,v) \in \mathcal{E}^a \). 
These variables jointly encode both {\bf function placement} and {\bf data routing} over the cloud-augmented graph. Concretely,  if \( u \) 
is a computation node, then \( f^k_{uv} = 1 \) models the placement of the function that produces commodity \( k \) at the computation node $u$. 
If both \( u \) and \( v \) are communication 
nodes, then \( f^k_{uv} = 1 \) denotes that commodity \( k \) is routed along the communication link \( (u,v) \).

The other key decision variables in the addressed service optimization problem are the {\bf resource allocation variables}, which are introduced in Section \ref{sec:queue_delay_modeling}.


\subsection{Computational Architecture for AI Workloads}
In AI-focused server environments, computation primarily occurs within containers and virtual machines running on dedicated hardware. 
Containers are favored because they offer strong isolation, near-native performance, and fine-grained resource control. 
A single server $u\in\V$ may host multiple containers,   each serving a different application component.
Whenever two commodities $k, l \in \mathcal{K}$ satisfy $l \in \mathcal{X}(k)$, we interpret $l$ as an input commodity that, through a service function executed in a container, contributes to the production of commodity $k$.
In our model,  each container processes exactly one commodity at a time, while the underlying operating system multiplexes containers via time-sharing.


Understanding how computational tasks are executed within containers is crucial for accurately modeling delays, as these delays vary significantly depending on the type of workload. In particular, it is essential to distinguish between CPU-bound and GPU-bound computations, which exhibit fundamentally different execution and scheduling behaviors.


CPU-based workloads can support multiple concurrent tasks in memory by leveraging fast context switching, which enables efficient time-sharing. This capability allows the operating system to define and guarantee processing quotas for each commodity, ensuring fair resource allocation and predictable performance. As a result, queuing delays in CPU-bound scenarios are primarily a consequence of time-sharing. 

Conversely, GPU computation, especially in AI applications, operates under a different paradigm. Specifcially, modern AI workloads executed on GPUs load large deep-learning models that often saturate VRAM \cite{hoffman2022,kaplan2020}.
Unlike CPU scheduling, switching between different AI models 
would require offloading and reloading model data into VRAM, introducing substantial I/O overhead and inefficiencies. As a result, a GPU 
effectively runs one model at a time, making its delay dynamics very different from those of a CPU.

Abstracting beyond the CPU/GPU dichotomy, we identify two computation-delay regimes:  i) guaranteed-quota regime where every task receives a dedicated slice of the resource, so its sojourn time is independent of other tasks on the same node;  ii) shared-resource regime where multiple tasks contend for a common resource pool, and each task’s delay depends on the aggregate load.


These regimes necessitate two distinct queuing models. Accordingly, we choose the appropriate queue class based on the architecture and physical characteristics of the node or link $(u,v)$ to accurately reflect the corresponding delay behavior. The formal delay expressions for both regimes are provided in the next section.

\section{Queue-Based Delay Modeling}
\label{sec:queue_delay_modeling}

In this section, we aim to develop a model for estimating the expected latency experienced by a commodity as it traverses the network. This latency comprises the waiting and processing delays encountered by requests across the network. 
To model these delays, we represent each link $(u,v) \in \Edges^a$ with one or more queues $Q_{uv}$. In the following, we assume that each queue possesses a single server, is non-preemptive, and is work-conserving, with a \gls{FCFS} scheduling policy.

We begin by characterizing the total average rate that may traverse link $(u,v)$, followed by two alternative models for the service rate, and finally derive the expected delay experienced by requests crossing $(u,v)$. Specifically, we assume that requests for a global service $\phi$ arrive according to a Poisson process with rate $\Lambda^\phi$. Each service $\phi$ is modeled as a directed acyclic graph (DAG) of commodities. 
The request arrival rate for all commodities $k \in \phi$ is given by the service request rate $\Lambda^\phi$. 
The {\bf request arrival rate} of commodity $k$ at link $(u,v)$ is hence given by $\lambda^k_{uv} = f^k_{uv} \Lambda^{\phi(k)}$, where recall that $\phi(k)$ denotes the service associated with commodity $k$. 
Accordingly, the total request arrival rate on link $(u,v)$ is given by $\lambda_{uv} = \sum_{k \in \mathcal{K}} \lambda^k_{uv}$.

\begin{figure}[t]
    \centering
    \begin{subfigure}[b]{0.97\linewidth}
        \centering
        \setlength{\fboxsep}{0pt}
        \fbox{\includegraphics[width=\linewidth]{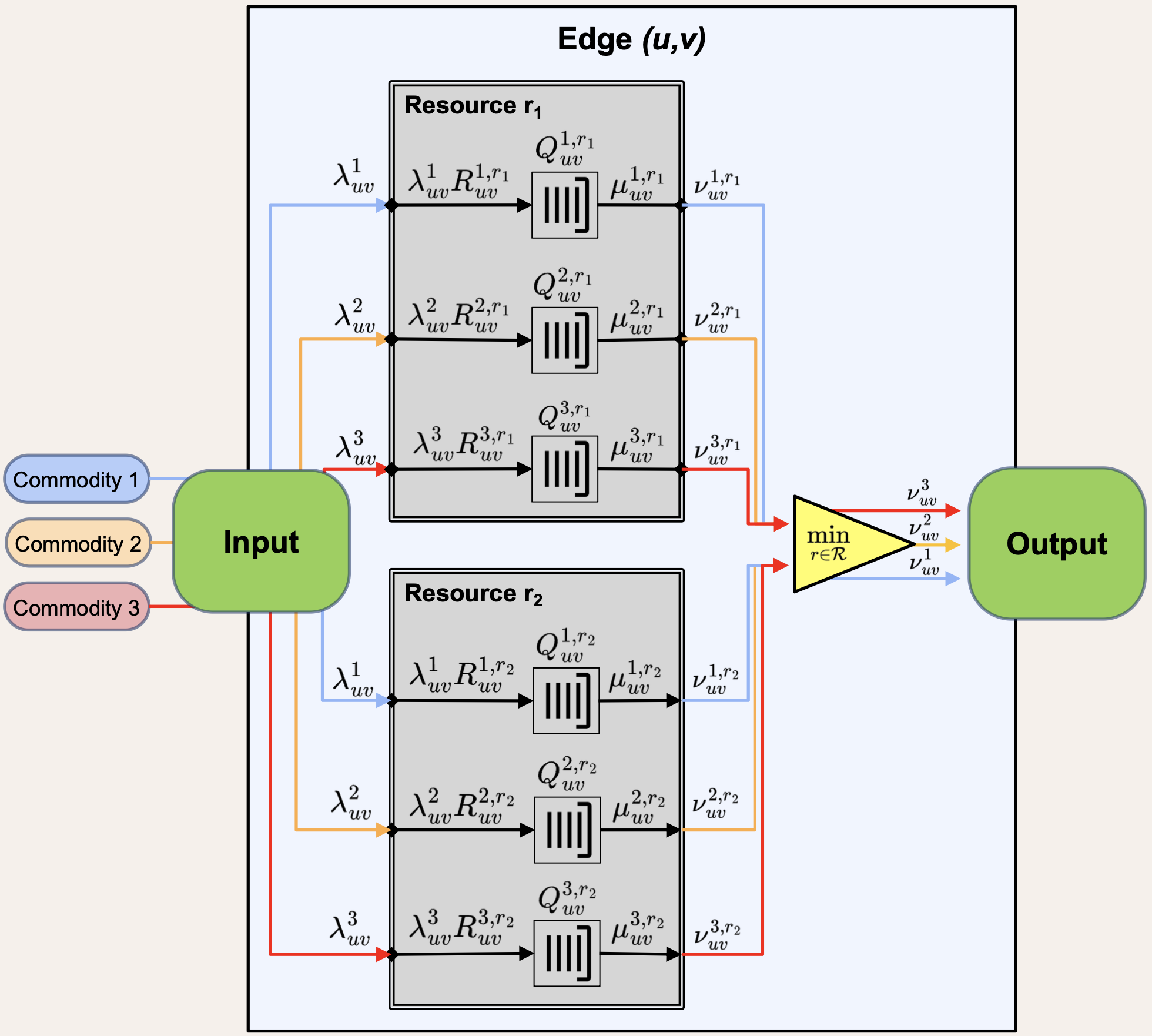}}
        \caption{Guaranteed-Resource model. Every incoming request has its own commodity-dedicated queue inside the system. Each queue system serves only one commodity. The system time experienced by one commodity does not influence other commodities.}
        \label{fig:model_guaranteed_resources}
        \vspace{0.3cm}
    \end{subfigure}
    
    \begin{subfigure}[b]{0.97\linewidth}
        \centering
        \setlength{\fboxsep}{0pt}
        \fbox{\includegraphics[width=\linewidth]{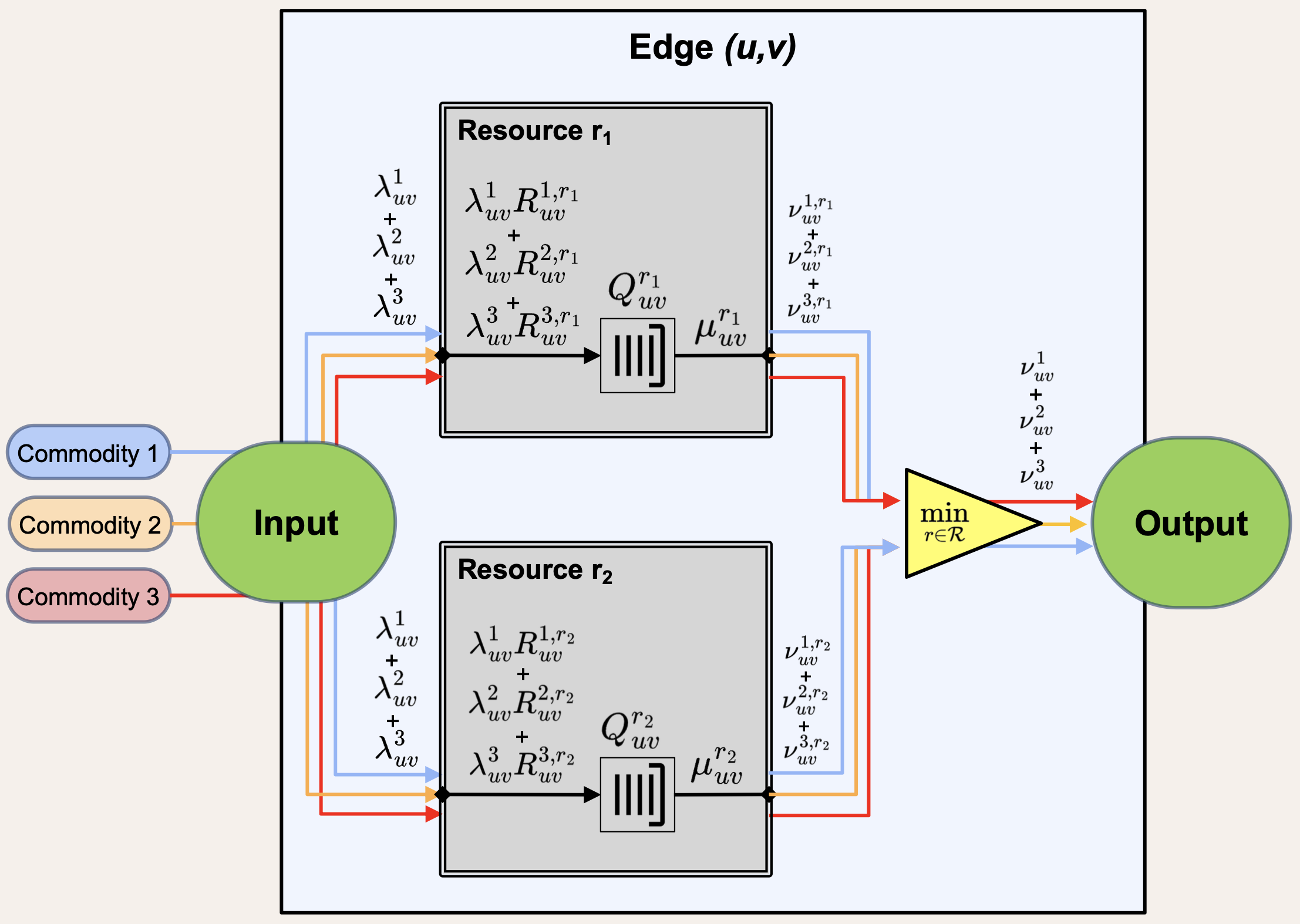}}
        \caption{Shared-Resource model. All requests flow inside the same queue, shared across commodities. The service requirements vary by commodity, resulting in a shared processing resource where delays depend on the combined workload. The system time experienced by one commodity is influenced and depends from other commodity using the same shared resource.}
        \label{fig:model_shared_resources}
    \end{subfigure}
    \caption{Flow rates for the Shared and Guaranteed Resource models.}
    \label{fig:model_resources}
\end{figure}

In the following subsections, we present two strategies for modeling {\bf resource allocation and associated delays}, illustrated in Fig.~\ref{fig:model_resources}, and provide a detailed description of each.

\subsection{Guaranteed-Resource model}
In this first approach, we assume that resources are exclusively allocated to each container processing a commodity, provided that the total allocated resources remain within the available capacity, as illustrated in Fig.~\ref{fig:model_guaranteed_resources}. This is referred to as the {\em Guaranteed-Resource} (GR) model.
Under this model, multiple processing containers can coexist on the same server without interference or contention, since each is assigned a dedicated portion of the available resources. This setup enables us to model processing using distinct and independent queues, each dedicated to a single commodity.
This model is particularly suitable for cases such as:
\begin{itemize}
    \item CPU processing within containerized environments, where each container is allocated a guaranteed CPU time slice, ensuring that its processing delays are independent of other containers;
    \item Network slices on communication infrastructure, where pre-provisioned network slices with fixed bandwidth allocations, where resources are reserved and isolated for specific services or tenants;
    \item Dedicated virtual machines or dedicated bare-metal servers, where computational resources are fully assigned and not shared dynamically with other workloads.
\end{itemize}

The set of links following the GR model is indicated as $\Edges^{a, \GR} \subseteq \Edges^a$. We define the decision variables that govern the {\bf resource service rate} at which $(u,v) \in \Edges^{a, \GR}$ serves commodity $k$ using resource $r$ as $\mu^{k,r}_{uv}$. 
Depending on the nature of resource $r$, it may be quantified in e.g., bits per second, operations per second, 
CPU/GPU cycles per second, etc. 

The demand for 
resource $r \in \R$ by commodity $k$ on link $(u,v)$ is denoted by $R^{k,r}_{uv}$ and is referred to as the {\bf resource-$r$ requirement} (or workload) of $k$ on $(u,v)$. Depending on the nature of resource $r$, it may be quantified in units of e.g., bits per request, operations per request, CPU/GPU cycles per request, etc. 

We can then derive the {\bf request service rate} of commodity $k$ on $(u,v)$, i.e., the rate at which $(u,v)$ serves requests of commodity $k$, as $\nu^k_{uv} = \min_{r\in\R}{\frac{\mu^{k,r}_{uv}}{R^{k,r}_{uv}}}$, measured in requests/s. The minimization over $r$ is needed because commodity $k$ may require service from multiple resources, and the overall output rate is constrained by the slowest rate among all resources, as shown in Fig. \ref{fig:model_guaranteed_resources}.
The service time for requests of commodity $k$ on link $(u,v)$ is modeled as a negative exponential distribution with rate $\nu^k_{uv}$. 
Finally, given that the arrival process for requests of commodity $k$ on $(u,v)$ follows a Poisson distribution with parameter $\lambda^k_{uv}$. 
we can model link $(u,v)$ as a set of {\bf M/M/1} queues $\{Q^{k,r}_{uv}\}$, with a separate queue for each commodity-resource pair $(k,r)$.

\subsection{Shared-Resource model}
In contrast to the GR model, the second approach assumes that computation relies on a single shared resource that is concurrently used by multiple commodities, as illustrated in Fig.~\ref{fig:model_shared_resources}. This is referred to as the {\em Shared-Resource} (SR) 
model. In this setting, resources are not partitioned among containers; instead, all active commodities compete for the same processing resources. As a result, the sojourn time of any given commodity is affected by the collective load imposed by all other commodities sharing the same resource.
This model is particularly suitable for environments such as:
\begin{itemize}
    \item GPU processing for heavy AI workloads, where multiple large AI models share the same GPU, which cannot efficiently interleave executions due to VRAM constraints, causing task delays to depend on all workloads on the GPU;
    \item Communication links without network slicing, where all traffic shares the same physical medium, leading to contention and interdependent delays among commodities routed on the link;
    \item Shared storage or I/O resources, where multiple processes access the same disk or network storage, leading to queuing delays affected by concurrent usage.
\end{itemize}


The set of $(u,v)$ following the SR model is indicated as $\Edges^{a, \SR} \subseteq \Edges^a$. 
In this case, we define the decisions variables that govern the {\bf resource service rate} at which $(u,v) \in \Edges^{a,\SR}$ serves {\em all} arriving commodities over resource $r$ as $\mu^{r}_{uv}$. 
The arrival process of the \SR model is the cumulative process obtained by summing the arrivals for all commodities routed on $(u,v)$, which, like in the \GR model, follows a Poisson distribution. 
The service time of requests for commodity $k$ is modeled as a negative exponential distribution with rate $\nu^k_{uv} = \min_{r\in\R}{\frac{\mu^{r}_{uv}}{R^{k,r}_{uv}}}$, named request service rate. In contrast to the \GR model, different commodities $j\neq k$ can be processed by the same queue. Since each commodity $k, j $ traversing $(u,v)$ requires different workloads $R^{k,r}_{uv}, R^{j,r}_{uv}$, the cumulative service time is general and doesn't follow an exponential distribution. Hence, given that the service request arrival process is Poisson and the individual commodity service times are exponential, a given network queue $Q^r_{uv}$ can be well-approximated by an {\bf M/G/1} model for large networks \cite{baiocchi2020}. Therefore, a link $(u,v)$ is modeled with a set of M/G/1 queues $\{Q^r_{uv}\}$, one for each resource, but shared among commodities.

A summary of the parameters and variables introduced in this section is given in Table \ref{tab:vars_description}.

\begin{table}[t]
\centering
\caption{Notation for the variables and parameters, alongside their measurement unit.}
\label{tab:vars_description}
\begin{tabular}{|c|l|l|}
\hline
\textbf{Symbol} & \textbf{Unit} & \textbf{Description} \\ \hline
\multicolumn{2}{|c|}{\textbf{Input Parameters}} & \\ \hline
$\Lambda^\phi$ & [request/s] & Request arrival rate of service $\phi$\\ \hline
$R^{k,r}_{uv}$ & [resource/request] & \makecell[l]{Resource-$r$ requirement of $k$ on $(u,v)$}\\ \hline
$L^k$ & [s] & Maximum admissible latency of $k$ \\ \hline
$M^r_{uv}$ & [resource/s] & Resource-$r$ capacity on $(u,v)$\\ \hline
$c^r_{uv}$ & [cost/resource/s] & \makecell[l]{Cost per unit of rate of operating $r$ \\ on $(u,v)$} \\ \hline
\multicolumn{2}{|c|}{\textbf{Decision Variables}} & \\ \hline
$f^k_{uv}$ & [Boolean] & Flow of commodity $k$ on edge $(u,v)$ \\ \hline
$\mu^r_{uv}$ & [resource/s] & \makecell[l]{Resource-$r$ service rate on $(u,v)$ \\ following \SR model} \\ \hline
$\mu^{k,r}_{uv}$ & [resource/s] & \makecell[l]{Resource-$r$ service rate of $k$ on $(u,v)$ \\ following \GR model} \\ \hline
 \multicolumn{3}{|l|}{\phantom{aaaa}\textbf{Auxiliary Variables} (resulting from decision variables)} \\ \hline
$\lambda^k_{uv}$ & [request/s] & Request arrival rate of $k$ on $(u,v)$ \\ \hline
$\lambda_{uv}$ & [request/s] & Request arrival rate on $(u,v)$ \\ \hline
$\nu^k_{uv}$ & [request/s] & Request service rate of $k$ on $(u,v)$ \\ \hline
$d^k_{uv}$ & [s] & Latency experienced by $k$ on $(u,v)$ \\ \hline
$l^k$ & [s] & Production-to-consumption latency of $k$ \\ \hline
$l^k_T$ & [s] & Total cumulative latency of $k$\\ \hline
$\rho^r_{uv}$ & $[0,1] \in \mathbb{R}$ & \makecell[l]{Utilization factor of resource $r$ on $(u,v)$ \\ following the \SR model}\\ \hline
$\rho^{k,r}_{uv}$ & $[0,1] \in \mathbb{R}$ & \makecell[l]{Utilization factor related to $k$ of resource \\ $r$ on $(u,v)$ following the \GR model}\\ \hline
$F_{uv}$ & [Boolean] & Activation of edge $(u,v)$\\ \hline
\end{tabular}
\end{table}

\subsection{Delay modeling}

We now analyze the delay associated with a generic queue in the system. Let $D^{k,r}_{uv}$ be a random variable expressing the sojourn time (queue time plus processing time) of commodity $k \in \K$ requiring resource-$r$ traversing link $(u,v) \in \Edges^a$.
Following the proposed models, we investigate the sojourn time $D^{k,r}_{uv}$ in four distinct cases, distinguishing whether $(u,v)$ is a communication or computation link, and whether it follows the \GR or \SR model. We begin by analyzing communication links the \GR and \SR models. Then, we generalize the obtained formulation to analyze the computation links within the \GR and \SR models.



\textbf{1) Communication links, \GR model.} 
Let $(u,v) \in \Edges^a$ denote a communication link modeled under the \GR framework. We model communication links with a single resource type $r\in\R$, which could be transmitted bits. Accordingly, $\mu^{k,r}_{uv}, R^{k,r}_{uv}, Q^{k,r}_{uv}, D^{k,r}_{uv}$ become independent of $r$, and we simplify the notation by writing them as $\mu^k_{uv}, R^k_{uv}, Q^{k}_{uv}, D^k_{uv}$, respectively.
As previously stated, a queue $Q^{k}_{uv}$ follows an M/M/1 model and a well-known closed-form solution exists for the sojourn time, given by the inverse of the difference between the request service rate and the request arrival rate. In our model, this is expressed as:
\begin{equation}
\label{eq:comm_mm1}
    \E[D^k_{uv}] = \frac{1}{\nu^k_{uv}-\lambda^k_{uv}} = \frac{R^k_{uv}}{\mu^k_{uv} - f^k_{uv} \Lambda^{\phi(k)} R^k_{uv}}
\end{equation}
where we applied the definitions of $\nu^k_{uv}$ and $\lambda^k_{uv}$.

\textbf{2) Communication links, \SR model.} 
Let $(u,v) \in \Edges^a$ be a communication link modeled under the \SR model.

We model communication links with a single resource type $r\in\R$. Accordingly $\mu^{r}_{uv}, R^{k,r}_{uv}, Q^r_{uv}, D^{k,r}_{uv}$ become independent of $r$, and we simplify the notation by writing them as $\mu_{uv}, R^k_{uv}, Q_{uv}, D^k_{uv}$, respectively.
In the \SR model, $Q_{uv}$ follows an M/G/1 model. We start by explicitly defining $D^k_{uv}$ as the sum of two random variables: $W_{uv}$, representing the queuing time, and $X^k_{uv}$, representing the service time of requests of commodity $k$ on queue $Q_{uv}$.
Our goal is to define and model the expected sojourn time $\E[D^k_{uv}]$:
\begin{equation}
    \E[D^k_{uv}] = \E[W_{uv}] + \E[X^k_{uv}]
\end{equation}
We note that the service time depends on the workload requested by $k$, while the waiting time is independent of the incoming commodities under the \gls{FCFS} scheduling policy.

Regarding the service time $\E[X^k_{uv}]$, since we have assumed exponential service times given a single commodity, 
we have $X^k_{uv} \sim f_{X^k_{uv}}(x) = \text{Exp}\left(\nu^k_{uv}\right)$. Therefore we have:
\begin{align}
    & \E[X^k_{uv}] = \frac{1}{\nu^k_{uv}} = \frac{R^k_{uv}}{\mu_{uv}} \\
    \label{eq:esp_Xkuv_squared}
    & \E[(X^k_{uv})^2] = 2\left(\frac{1}{\nu^k_{uv}}\right)^2 = 2\left(\frac{R^k_{uv}}{\mu_{uv}}\right)^2
\end{align}

To compute the waiting time $\E[W_{uv}]$, we apply the \textit{Pollaczek--Khinchine formula}~\cite{baiocchi2020, pollaczek_khinchine} for the expected waiting time in an M/G/1 queue:
\begin{equation}
\label{pollaczek}
    \E[W_{uv}] = \frac{\lambda_{uv}\E[X^2_{uv}]}{2(1-\rho_{uv})}
\end{equation}
where the random variable $X_{uv}$ represents the service time of a generic request on queue $Q_{uv}$, regardless of the specific commodity being served.
To complete \eqref{pollaczek}, we first compute the second moment of the service time, $\E[X^2_{uv}]$, and then the utilization factor, $\rho_{uv}$.
The probability density function and the expected value of $X_{uv}$ is given by:
\begin{align}
    \label{eq:X_uv_pdf}
    X_{uv} \sim f_{X_{uv}}(x) & = \sum_{k \in \K} \frac{\lambda^k_{uv}}{\lambda_{uv}} f_{X^k_{uv}}(x) \\
    \label{eq:X_uv_exp}
    \E[X_{uv}] & = \sum_{k\in\K} \frac{\lambda^k_{uv}}{\lambda_{uv}} \E[X^k_{uv}]
\end{align}
where the term $\frac{\lambda^k_{uv}}{\lambda_{uv}}$ represents the probability that a randomly selected request in queue $Q_{uv}$ belongs to commodity $k$, thereby defining the contribution of $f_{X^k_{uv}}(x)$ to the overall service time distribution. To compute the second moment of the service time, we proceed as follows:

\begin{align}
    \label{eq:X_uv_exp_squared}
    \E[(X_{uv})^2] &= 
    \int_0^{+\infty} x^2 \cdot f_{X_{uv}}(x) dx = \\
     &= \sum_{k\in\K} \frac{\lambda^k_{uv}}{\lambda_{uv}} \int_0^{+\infty} x^2 \cdot f_{X^k_{uv}}(x) = \nonumber \\
    &= 
    \sum_{k\in\K} \frac{\lambda^k_{uv}}{\lambda_{uv}} \E[(X^k_{uv})^2] \nonumber
\end{align}
Now, using \eqref{eq:esp_Xkuv_squared} and the definition of $\lambda^k_{uv}$ we have:
\begin{equation}
\label{nz:EX_squared}
    \E[(X_{uv})^2] = 2 \sum_{k \in \K} \frac{f^k_{uv}\Lambda^{\phi(k)}}{\lambda_{uv}} \left(\frac{R^k_{uv}}{\mu_{uv}}\right)^2
\end{equation}
Note that \eqref{nz:EX_squared} captures the contribution of each individual commodity to the second moment of the service time, thereby enabling accurate modeling of queuing behavior. This formulation, which incorporates the flow variables \( f^k_{uv} \), reflects whether a commodity is actively routed over the link and accounts for its precise impact on the congestion.


We now turn to the {\em utilization factor} $\rho_{uv}$, which, along with $\E[X^2_{uv}]$, completes the Pollaczek-Khinchine expression for the queuing time. As discussed earlier, since each communication link is associated with a single resource type $r \in R$, we simplify the notation by writing $\rho^r_{uv}$ as $\rho_{uv}$.

To compute $\rho_{uv}$, we begin by defining the {\em per-commodity utilization} $\rho^k_{uv} = \lambda^k_{uv} / \nu^k_{uv}$, which represents the fraction of time queue $Q_{uv}$ is busy serving commodity $k$. The total utilization of the queue is then the sum over all commodities:
\[
\rho_{uv} = \sum_{k \in \K} \rho^k_{uv}.
\]
Substituting the expressions for $\lambda^k_{uv}$ and $\nu^k_{uv}$ yields:
\begin{equation}
\label{rho}
\rho_{uv} = \sum_{k \in \K} \frac{\lambda^k_{uv}}{\nu^k_{uv}} = \frac{1}{\mu_{uv}} \sum_{k \in \K} f^k_{uv} \Lambda^{\phi(k)} R^k_{uv}.
\end{equation}
This expression offers a clear physical interpretation: $\rho_{uv}$ represents the ratio between the {\em total workload} imposed on the link by all active commodities and the {\em allocated resources}, expressed by the service rate $\mu_{uv}$. It captures how congestion builds up based on both the intensity of incoming traffic and the per-commodity resource demands.

Combining (\ref{nz:EX_squared}) and \eqref{rho} into \eqref{pollaczek}, the final expected sojourn time of commodity $k$ on queue $Q_{uv}$ is:
\begin{align}
\label{eq:comm_mg1}
    \E[D^k_{uv}] &= \E[W_{uv}] + \E[X^k_{uv}] \nonumber \\
    &= \frac{\sum_{j \in \K} f^j_{uv} \Lambda^{\phi(j)} (R^j_{uv})^2}{\mu_{uv}\left(\mu_{uv} - \sum_{j\in\K} f^j_{uv}\Lambda^{\phi(j)} R^j_{uv}\right)} + \frac{R^k_{uv}}{\mu_{uv}}
\end{align}
The derived formula represents the expected sojourn time $\E[D^k_{uv}]$ of a commodity $k$ being routed into queue communication link $(u,v)$ following the \SR model.

\textbf{3) Computation links, \GR model.}
Let $(u,v) \in \Edges^a$ be a computation link, where $v = p_i$, modeled under the \GR framework. We now derive the sojourn time formulation for computation links—and more generally, for commodities that require multiple resource types on the same link.
To begin, we express the expected sojourn time for a commodity $k$ with respect to each individual resource $r \in \R$ using the M/M/1 delay model:
\begin{align}
\label{eq:soj_proc_mm1}
\E[D^{k,r}_{uv}] &= \frac{R^{k,r}_{uv}}{\mu^{k,r}_{uv} - f^k_{uv} \Lambda^{\phi(k)} R^{k,r}_{uv}}
\end{align}

Since a commodity may require service from multiple resources on the same node, its overall sojourn time is determined by the slowest (i.e., most congested) resource. Therefore, the total expected sojourn time is lower bounded by:
\begin{equation}
\label{eq:maximization_gr}
    \E\left[D^k_{uv}\right] =
    \E\left[\max_{r \in \R} D^{k,r}_{uv}\right] \geq
    \max_{r \in \R} \E\left[D^{k,r}_{uv}\right]
\end{equation}
where the inequality follows from Jensen’s inequality~\cite{jensenineq}, given that the max function is convex.

\textbf{4) Computation links, \SR model.}
Let $(u,v) \in \Edges^a$ be a computation link, with $v=p_i$, modeled following the \SR model. In line with the approach used for the \GR case, we express the average sojourn time of a \SR link for each resource $r\in\R$ as follows:
\begin{align}
\label{eq:soj_proc_mg1}
\E[D^{k,r}_{uv}] &= \frac{\sum_{j \in \K} f^j_{uv} \Lambda^{\phi(j)} (R^{j,r}_{uv})^2}{\mu^r_{uv}\left(\mu^r_{uv} - \sum_{j\in\K} f^j_{uv}\Lambda^{\phi(j)} R^{j,r}_{uv}\right)} + \frac{R^{k,r}_{uv}}{\mu^r_{uv}}
\end{align}
To obtain an overall estimate of the sojourn time, we again apply the maximization over resources:
\begin{equation}
\label{eq:maximization_sr}
    \E\left[D^k_{uv}\right] \geq
    \max_{r \in \R} \E\left[D^{k,r}_{uv}\right]
\end{equation}

\textbf{Delay Constraints and Optimization Challenges:}
Together, Equations \eqref{eq:comm_mm1}, \eqref{eq:comm_mg1}, \eqref{eq:maximization_gr} and \eqref{eq:maximization_sr} give different formulations of the sojourn time $D^{k,r}_{uv}$ depending on whether $(u,v)$ is a communication or computation link and whether it follows the \GR or the \SR model. We aim to constrain the various formulations of the sojourn time within a maximum limit by summing the sojourn times across all queues that a commodity traverses. However, the expression of the sojourn time for the \GR model is non-convex both in $f^k_{uv}, \mu^{r}_{uv}$ and $\mu^{k,r}_{uv}$, making direct optimization challenging. In Sec. \ref{sec:prob_formulation} we formulate the problem and then, in Sec. \ref{sec:alg_sol}, we will address the optimization by introducing suitable approximations and developing efficient approximation algorithms.

\section{Problem Formulation}
\label{sec:prob_formulation}
In this section, we describe the problem formulation. Our goal is to minimize a cost function that depends on service placement, routing, and resource allocation, subject to latency and capacity constraints. We formulate problem $\mathcal{P}$ as follows:
\begin{align}
    && & \min_{F_{uv}, \mu^r_{uv}, \mu^{k,r}_{uv}} \sum_{(u,v) \in \mathcal{E}^{a,\SR}} \sum_{r\in\R} F_{uv} \mu^r_{uv} c^r_{uv} + \span\span\quad(\mathcal{P})\nonumber \\
    && & \quad\quad\quad\;\;\;+\sum_{(u,v)\in\Edges^{a,\GR}}\sum_{k\in\K}\sum_{r\in\R} F_{uv}\mu^{k,r}_{uv}c^r_{uv}\,\textnormal{, s.t.:} \span\span\nonumber\\
    && & (a_1)\; \sum_{\mathclap{v \in N^-(u)}} f^k_{vu} = \sum_{\mathclap{v \in N^+(u)}} f^k_{uv}
            & & \forall u \in \V, k \in \K \nonumber\\
    && & (a_2)\; f^k_{pu} = f^l_{up}
            & & \forall (u,p) \in \Edges^a, k \in \K, l \in \X(k) \nonumber\\
    && & (a_3)\; f^k_{su} = 0
            & & \forall k \in \K, u \neq s(k) \nonumber\\
    && & (a_4)\; f^k_{ud} = 1
            & & \forall k \in \K^d, u = d(k) \nonumber\\
    && & (a_5)\; F_{uv} \geq f^k_{uv}
            & & \forall k \in \K, (u,v) \in \mathcal{E}^a \nonumber\\
    && & (b_1)\; d^k_{uv} \geq \E[D^{k,r}_{uv}(\bm{f},\bm{\mu})]
        & & \forall (u,v)\in\Edges^a, k \in \K, r \in \R \nonumber\\
    && & (b_2)\; l^k = \sum_{\mathclap{(u,v) \in \mathcal{E}^a}} f^k_{uv} d^k_{uv}
        & & \forall k \in \K \nonumber\\
    && & (b_3)\; l^k_T \geq l^k + l^j_T
        & & \forall k \in \K \setminus \K^s, j \in \X(k) \nonumber\\
    && & (b_4)\; l^k_T = l^k
        & & \forall k \in \K^s \nonumber\\
    && & (b_5)\; l^k_T \leq L^k
        & & \forall k \in \K^d \nonumber\\
    && & (c_1)\, \mu^r_{uv} >\, \mathclap{\sum_{k\in\K}}\;\;\, f^k_{uv} \Lambda^{\phi(k)} R^{k,r}_{uv}
            \; \forall (u,v) \in \mathcal{E}^{a,\SR}, r\in\R \span \span \nonumber\\
    && & (c_2)\, \mu^{k,r}_{uv} > f^k_{uv} \Lambda^{\phi(k)} R^{k,r}_{uv}
            && \forall (u,v) \in \mathcal{E}^{a,\GR}, k\in\K, r\in\R  \nonumber\\
    && & (e_1)\; \mu^r_{uv} \leq M^r_{uv}
            && \forall (u,v) \in \mathcal{E}^{a,\SR}, r \in \R \nonumber\\
    && & (e_2)\; \sum_{k\in\K}\mu^{k,r}_{uv} \leq M^r_{uv}
            && \forall (u,v) \in \mathcal{E}^{a,\GR}, r \in \R \nonumber\\
    && & (e_3)\; f^k_{uv} \in \{0,1\}
        && \forall (u,v) \in \mathcal{E}^a, k \in \K \nonumber\\
    && & (e_4)\; \mu^{r}_{uv}, \mu^{k,r}_{uv} \geq 0
            & & \forall (u,v) \in \mathcal{E}^{a}, k \in \K, r \in \R \nonumber
\end{align}

The objective function to minimize is the total operational cost. In our notation, $c^r_{uv}$ represents the cost per unit of resource-r service rate allocated on link $(u,v) \in \Edges^a$. The objective function is composed of two terms, accounting for the links that follow the \SR and \GR model respectively. The auxiliary variable $F_{uv}$ takes the value 1 if at least one commodity traverses link $(u,v)$, meaning the link needs to be activated, and 0 otherwise. Therefore, our cost model accounts for the links used in the augmented graph and the resource allocated to each of them. This cost model is particularly suitable to cloud on-demand services, like Amazon \gls{AWS} or similar, where costs are influenced by both the geographical location of processing and the computational resources assigned to the VMs executing the operations.

The variables to be optimized are the flow variables and resource allocation variables. The flow variables are represented as $f^k_{uv}, (u,v)\in\Edges^a, k\in\K$. The resource allocation variables are represented with resource-r service rates $\mu^{k,r}_{uv}, (u,v)\in\Edges^{a,\GR}, k\in\K, r\in\R$ in case a link is modeled with the \GR model, requiring separate guaranteed resources for each commodity, and $\mu^r_{uv}, (u,v)\in\Edges^{a,\SR}, r\in\R$ in case a link is modeled with the \SR model, where a resource is shared across commodities.

Equations $(a_1)-(a_5)$ are flow conservation constraints. In particular, $(a_1)$ ensures that the flow of a given commodity remains constant as it traverses a network link between two nodes, and $(a_2)$ ensures that the flow consumed as input to create a new commodity is fully utilized in producing the new commodity for all its input components. This constraint preserves the \gls{DAG} relationship between commodities. In the equation, we used the notation $f^k_{up}$ indicating the flow passing through network node $u \in \V$ and its associated computation node in the augmented graph. Equation $(a_3)$ ensures that no source node generates flow unless it is the node designated to generate the flow as specified by the problem's input. Similarly, $(a_4)$ requires that all the flow must be directed to the designated destination node. Finally, constraint $(a_5)$ deals with the $F_{uv}$ indicator variable, assuming value 1 if link $(u,v)$ is utilized, 0 otherwise.

Equations $(b_1)-(b_5)$ are delay constraints. Constraint $(b_1)$ performs a maximization and bounds the auxiliary variable $d^k_{uv}$ to the sojourn time of a request of commodity $k$ on link $(u,v)$. We indicate with $\bm{f}, \bm{\mu}$ the vectors containing $f^k_{uv}, \mu^r_{uv}$ variables. In the constraint, we make explicit the dependence of the average delay on $\bm{f}, \bm{\mu}$. We remark that the formulation of the average delay differs depending on whether $(u,v)$ is a processing or a communication link, and on whether it follows the \GR or \SR model. Constraint, $(b_2)$ defines $l^k$ as the expected production-to-consumption delay experienced by commodity $k$ as the summation of the sojourn times across all the queues it traverses. Equation $(b_3)$ defines $l^l_T$ which is the total cumulative latency of commodity $k$, taking into consideration the delay of input commodities $j \in \X(k)$. Then, equation $(b_4)$ sets the total cumulative latency for source commodities to $l^k$, as they are not generated by any other commodity. Constraint $(b_5)$ bounds the total cumulative latency to the maximum allowable end-to-end latency $L^k$, which is provided as an input to the problem.


Continuing, $(c_1)-(c_2)$ are queue stability bounds ensuring that the resource-r service rate is always greater than the incoming workload arrival rate. 
Finally, $(e_1) - (e_2)$ bound the allocated rates to be not greater than the system capacity, and $(e_3)-(e_4)$ define the domain of the variables.

We have presented the problem formulation along with all relevant constraints. Depending on whether an edge is modeled using the \GR or \SR approach, slightly different constraints are applied. The choice between the two ultimately depends on the specific problem setting and which model better reflects the physical characteristics of the underlying resource.
In the case of the \GR model for example, we have guaranteed and reserved resources per commodity, there is no contention among different commodities, and the delay experienced by one does not affect the delay of others. In contrast, the \SR model assumes that the delays incurred by one commodity affect the delay of all other commodities sharing the same resource. In this case, all commodities routed over the same link share the medium, and their delays are mutually dependent.
Ultimately, the decision to model a link using the GR or SR model depends on the specific nature of the link and the characteristics of the environment. The proposed model is general and we presented different formulations of the average commodity sojourn time on a link following either the \GR or \SR model. Ultimately, the operator, who has knowledge of the application, the environment, and the network, must decide whether to model a link using the SR or GR model based on which approach best fits the scenario.

The presented formulation of problem $\mathcal{P}$ cannot be directly optimized for two main reasons. First, the integer domain of $f^k_{uv}$ makes $\mathcal{P}$ a mixed integer program, which is NP-hard. Secondly, the calculation of $\E[D^k_{uv}]$ is non-convex in the \SR model, as in Equations \eqref{eq:comm_mg1} and \eqref{eq:soj_proc_mg1}, due to the interdependent delays, even considering the linear relaxation of $\mathcal{P}$. Thus, in the next section, we introduce an algorithmic framework designed to find an approximated solution of the problem.

\section{Algorithmic Solution}
\label{sec:alg_sol}
In this section, we present the key steps involved in addressing the optimization problem. 
The main issue in approaching the problem are Equations \eqref{eq:comm_mg1} and \eqref{eq:soj_proc_mg1}, whose non-linearity and non-convexity make direct optimization intractable.

To overcome this, we first introduce a tractable upper bound to the sojourn time by bounding the utilization factor of a queue.
This step leads to a \emph{bi-convex formulation}: the problem is not jointly convex, but it is convex in each block of variables when the other is fixed. We therefore decompose it into two convex sub-problems and solve them iteratively, optimizing $f^k_{uv}$ and $\mu^r_{uv}$ in turn.
Finally, we propose an approximation algorithm that alternately optimizes these sub-problems.


\subsection{Delay upper bound}
\label{sec:eps_bound}
To make the derivation of the upper bound explicit, we apply the well-known $\eps$-safety method to bound the utilization factor, obtaining a delay formulation that can be incorporated into our algorithm. Specifically, we bound the utilization factor of queue
$Q^r_{uv}$ as follows:
\begin{equation}
\label{rho_bound}
    \rho^r_{uv} \leq 1-\eps^r_{uv}, \;\; \eps^r_{uv} \in [0,1], \forall r\in\R
\end{equation}
for a sufficiently small $\eps^r_{uv}$. This allows us to add an $\eps$-safety margin on constraint $(c_2)$ as follows:
\begin{equation}
\label{eps_margin}
    (1-\eps^r_{uv})\mu^r_{uv} \geq \sum_{k\in\K} f^k_{uv}\Lambda^{\phi(k)} R^{k,r}_{uv}
\end{equation}

It is well known that this method ensures that the system operates below its capacity avoiding potential instability. We now find the sojourn time when the system is at its maximum permitted load, applying equation \eqref{eps_margin} into \eqref{eq:comm_mg1} (same applies for \eqref{eq:soj_proc_mg1}). In this way we find an upper bound for the sojourn time, defined as $\E[\bar{D}^{k,r}_{uv}(\bm{f},\bm{\mu},\bm{\eps})]$:
\begin{align}
\E[D^{k,r}_{uv}(\bm{f},\bm{\mu})] &\leq \E[\bar{D}^{k,r}_{uv}(\bm{f},\bm{\mu},\bm{\eps})] = \\ 
&= \frac{\sum_{j\in\K} f^j_{uv} \Lambda^{\phi(k)} (R^{j,r}_{uv})^2}{\eps^r_{uv} (\mu^r_{uv})^2} + \frac{R^{k,r}_{uv}}{\mu^r_{uv}} \nonumber
\label{eq:dk_approx}
\end{align}
We observe that the previous formulation is convex in $f^k_{uv}$ if we fix $\mu^r_{uv}$ and viceversa is convex in $\mu^r_{uv}$ if we fix $f^k_{uv}$. 
In the optimization problem, we use this upper bound as the delay metric to be minimized. In other words, the system is optimized under its maximum load. This approach is inherently suboptimal, since the solution may over-allocate resources in order to satisfy the latency constraints. However, the parameter $\eps^r_{uv}$ is optimized iteratively by the approximation algorithm discussed in Section~\ref{sec:approx_algorithm}, which allows the solution to approach near-optimal performance.


\subsection{Problem decomposition and convexification}
In this section we decompose problem $\mathcal{P}$ into two sub-problems, namely \Pone$(\bm{\mu}, \bm{\eps}, i)$ and \Ptwo$(\bm{f}, \bm{\eps}, i)$. In \Pone~we optimize $\bm{f}$ variables and maintain $\bm{\mu}$ fixed, while in \Ptwo~we optimize $\bm{\mu}$ variables and maintain $\bm{f}$ fixed. The problems are used in the iterative algorithm presented in Section \ref{sec:approx_algorithm}, and $i$ indicates the iteration number. We indicate as $f^k_{uv}(i), \mu^r_{uv}(i), \mu^{k,r}_{uv}(i), \eps^r_{uv}(i)$ the values of the variables at iteration $i$. We formulate \Pone$(\bm{\mu}, \bm{\eps}, i)$ as follows:

\begin{align}
    && & \min_{F_{uv}} \sum_{(u,v) \in \mathcal{E}^{a,\SR}} \sum_{r\in\R} \mu^r_{uv}(i) F_{uv} c^r_{uv} + \span\span\quad(\mathcal{P}_1(\bm{\mu}, \bm{\eps}, i))\nonumber \\
    && & \quad+\sum_{(u,v)\in\Edges^{a,\GR}}\sum_{k\in\K}\sum_{r\in\R} \mu^{k,r}_{uv}(i) F_{uv} c^r_{uv}\,\textnormal{, s.t.:} \span\span\nonumber\\
    && & \frac{1}{1-\eps^r_{uv}(i)}\sum_{k \in \K} \hat{f}^k_{uv} \Lambda^{\phi(k)} R^{k,r}_{uv} \leq {\mu^r_{uv}(i)} 
    \span \span | \forall (u,v) \in \mathcal{E}^a, r\in\R \nonumber\\
    && & \sum_{\mathclap{v \in N^-(u)}} \hat{f}^k_{vu} =  \sum_{\mathclap{v \in N^+(u)}} \hat{f}^k_{uv} 
    && \forall u \in \V, k \in \K \nonumber\\
    && & \hat{f}^k_{pu} = \hat{f}^l_{up} 
    && \forall u \in \V, k \in \K, l \in \X(k) \nonumber\\
    && & \hat{f}^k_{su} = 0 
    && \forall k \in \K, u \neq s(k) \nonumber\\
    && & \hat{f}^k_{ud} = 1 
    && \forall k \in \K^d, u = d(k) \nonumber\\
    && & F_{uv} \geq \hat{f}^k_{uv} 
    && \forall k \in \K, (u,v) \in \mathcal{E}^a \nonumber\\
    && & \bar{d}^k_{uv} \geq \E\left[\bar{D}^{k,r}_{uv}(\bm{f},\bm{\mu}(i),\bm{\eps}(i))\right]
    \span \span \forall (u,v)\in\Edges^a, k\in\K, r\in\R \nonumber\\
    && & l^k = \sum_{\mathclap{(u,v) \in \mathcal{E}^a}} \hat{f}^k_{uv} \bar{d}^k_{uv}
    && \forall k \in \K  \nonumber\\
    && & l^k_T = l^k 
    && \forall k \in \K^s \nonumber\\
    && & l^k_T \geq l^k + l^j_T 
    && \forall k \in \K \setminus \K^s, j \in \X(k) \nonumber\\
    && & l^k_T \leq L^k 
    && \forall k \in \K^d \nonumber\\
    && & \hat{f}^k_{uv} \in [0,1] 
    && \forall (u,v) \in \mathcal{E}^a, k \in \K \nonumber
\end{align}
Here, we have incorporated the utilization factor bound from Sec.~\ref{sec:eps_bound} into the queue stability constraint, ensuring that the delay formulation remains tractable.
Also, we have introduced the variables $\hat{f}^k_{uv}$ which represent the continuous relaxation for the original problem.
Nevertheless, the problem remains non-convex: by expanding the definition of $l^k$, we obtain bilinear terms of the form $\hat{f}^k_{uv} \cdot \hat{f}^j_{uv}$. 
To handle this, we adopt the method described in \cite{scutari2017}, where bilinear terms $g(\hat{f}^k_{uv}, \hat{f}^j_{uv}) = \hat{f}^k_{uv} \cdot \hat{f}^j_{uv}$ are convexified by linearizing their concave component. 
Following \cite{scutari2017}, for any given $y_1, y_2 \in \mathbb{R}$, the concave part can be approximated using the values $\hat{f}^k_{uv}(i-1), \hat{f}^j_{uv}(i-1)$ from the previous iteration. 
This procedure ensures that at each iteration $i$, problem \Pone$(\bm{\mu}, \bm{\eps}, i)$ becomes convex in the flow variables $\hat{f}^k_{uv}$ and can therefore be solved efficiently.

We now formulate \Ptwo~as follows:
\begin{align}
    && & \min_{\mu^r_{uv}, \mu^{k,r}_{uv}} \sum_{(u,v) \in \mathcal{E}^{a,\SR}} \sum_{r\in\R} F_{uv}(i) \mu^r_{uv} c^r_{uv} + \span\span\quad(\mathcal{P}_2(\bm{f},\bm{\eps}, i))\nonumber \\
    && & \quad\quad+\sum_{(u,v)\in\Edges^{a,\GR}}\sum_{k\in\K}\sum_{r\in\R} F_{uv}(i) \mu^{k,r}_{uv} c^r_{uv}\,\textnormal{, s.t.:} \span\span\nonumber\\
    && & \mu^r_{uv} \geq \frac{\sum_{k \in \K} f^k_{uv}(i) \Lambda^{\phi(k)} R^{k,r}_{uv}}{1-\eps^r_{uv}(i)}
    &&  \forall (u,v) \in \mathcal{E}^a, r\in\R \nonumber\\
    && & \mu^{k,r}_{uv} \geq \frac{f^k_{uv}(i) \Lambda^{\phi(k)} R^{k,r}_{uv}}{1-\eps^r_{uv}(i)} 
    \span\span\;\;\;\;\;\; \forall (u,v) \in \mathcal{E}^a, k\in\K, r\in\R \nonumber\\
    && & \bar{d}^k_{uv} \geq \E\left[\bar{D}^{k,r}_{uv}(\bm{f}(i),\bm{\mu},\bm{\eps}(i))\right]
    \span \span | \forall (u,v)\in\Edges^a, k\in\K, r\in\R \nonumber\\
    && & l^k =  \sum_{\mathclap{(u,v) \in \mathcal{E}^a}} f^k_{uv}(i)\; \bar{d}^k_{uv}
    && \forall k \in \K \nonumber\\
    && & l^k_T = l^k 
    && \forall k \in \K^s \nonumber\\
    && & l^k_T \geq l^k + l^j_T 
    && \forall k \in \K \setminus \K^s, j \in \X(k) \nonumber\\
    && & l^k_T \leq L^k 
    && \forall k \in \K^d \nonumber\\
    && & \mu^r_{uv} \leq M^r_{uv}
    && \forall (u,v) \in \mathcal{E}^a, r\in\R \nonumber\\
    && & \sum_{k\in\K}\mu^{k,r}_{uv} \leq M^r_{uv}
    && \forall (u,v) \in \mathcal{E}^a, r\in\R \nonumber\\
    && & \mu^{r}_{uv}, \mu^{k,r}_{uv} \geq 0
    \span\span \forall (u,v) \in \mathcal{E}^{a}, k \in \K, r \in \R \nonumber
\end{align}
where we have used $F_{uv} = \max_{k \in \K} \hat{f}^k_{uv}$.
We note that \Ptwo~is convex for resource allocation variables $\mu^r_{uv}$.


\subsection{Approximation algorithm}
\label{sec:approx_algorithm}
We now define the SPARQ algorithm that optimizes \Pone~and \Ptwo~in an alternating and iterative manner, converging to a solution in the continuous domain of the original problem described in Section \ref{sec:prob_formulation}.
The algorithm is shown in Alg. \ref{alg} and it's an expansion of the NOVA algorithm presented in \cite{scutari2017}. In the algorithm, we indicate $\bm{\bar{f}}(i) = \text{\Pone}(\bm{\mu}, \bm{\eps}, i)$ to be the solution of \Pone~and $\bm{\bar{\mu}}(i) = \text{\Ptwo}(\bm{\bar{f}}, \bm{\eps}, i)$ to be the solution of \Ptwo. The steps of the algorithm are as follows:
\begin{algorithm}
\caption{Service Placement, Resource Allocation and Routing with Queue-Aware Delays (SPARQ)}
\label{alg}
\begin{algorithmic}[1]
    \STATE Initialize $\bm{\mu}(0) \leftarrow M, \bm{\eps}(0) \leftarrow \frac{3}{4}$
    \FOR{$i = 1$ to maximum iterations or convergence}
        \STATE $\bm{\bar{f}}(i) \leftarrow \textnormal{\Pone}(\bm{\mu}, \bm{\eps}, i-1)$
        \STATE $\bm{\bar{\mu}}(i) \leftarrow \textnormal{\Ptwo}(\bm{\bar{f}}, \bm{\eps}, i-1)$
        \STATE $\bm{\hat{f}}(i+1) \leftarrow \bm{\hat{f}}(i) + \gamma_i (\bm{\bar{f}}(i) - \bm{\hat{f}}(i))$
        \STATE $\bm{\mu}(i+1) \leftarrow \bm{\mu}(i) + \gamma_i (\bm{\bar{\mu}}(i) - \bm{\mu}(i))$ 
        \STATE Compute $\bm{{\rho}}(i) \textnormal{ with } \bm{\bar{f}}(i), \bm{\bar{\mu}}(i)$ following Eq. \eqref{rho}
        \STATE $\bm{\eps}(i+1) \leftarrow \gamma_i \bm{\eps}(i) + (1 - \gamma_i)(1 - \bm{{\rho}}(i))$
        \STATE $i \leftarrow i+1$
    \ENDFOR
    \STATE $\bm{f} \leftarrow$ Decomposition and Rounding IDAGO of $\bm{\hat{f}}(i+1)$
\end{algorithmic}
\end{algorithm}


The algorithm starts by initializing the resource allocation variables to their capacity $\bm{M}$, indicating the vector containing $M^r_{uv}$ values. It then proceeds by solving sub-problem \Pone~yielding the flow values $\bm{\bar{f}}(i)$. Next, sub-problem \Ptwo~is solved using $\bm{\bar{f}}(i)$ as input flow, resulting in $\bm{\bar{\mu}}(i)$, which represents the minimum service rate required to sustain the flow $\bm{\bar{f}}(i)$ under the latency constraints.
Then, we use a factor $\gamma_i$ to gradually update the variables. We use classical diminishing step-size rules for $\gamma_i$, which satisfy the conditions in \cite{scutari2017}. Such approach guarantees asymptotic convergence of the algorithm to stationary points of Problem $\mathcal{P}$.
Continuing, the algorithm computes the vector $\bm{{\rho}}(i)$, which contains the observed utilization factor $\rho^r_{uv}$ with the instantaneous $\bm{\bar{f}}(i), \bm{\bar{\mu}}(i)$ values, computed following Eq. \eqref{rho}. 
Then, the vector $\bm{\eps}(i)$ is updated as a function of the computed utilization $\bm{\rho}(i)$, and the procedure is repeated until convergence.
This update adaptively tunes the safety margins to reflect the current system load, thereby improving the tightness of the utilization factor and delay bound.

In general, the algorithm iterates by progressively selecting smaller values of $\bm{\mu}$, optimizing the flow at each step. With each iteration, it moves closer to the optimal solution, refining both the flow and the resource allocation variables.
We observe that the algorithm never assigns values to the variables $\bm{\mu}$ that exceed their respective upper limits $\bm{M}$: initially, $\bm{\mu}$ is set to their maximum allowable values, i.e., $\bm{\mu}(0) = \bm{M}$. In the subsequent steps, \Pone~is solved. Since the initial values of $\bm{\mu}(0)$ are at their upper bounds, the solution $\bm{\bar{f}}(i)$ is obtained using the maximum possible capacity of the system. Next, \Ptwo~is solved to determine the minimal values of $\bm{\bar{\mu}}(i)$ that are sufficient to support the flow $\bm{\bar{f}}(i)$. Importantly, the flow $\bm{\bar{f}}(i)$ was already supported by the maximum capacity found in the previous step. Thus, either the problem is unfeasible or the solution of \Ptwo~must produce $\bm{\bar{\mu}}(i)$ values that are no greater than the current $\bm{\mu}(i-1)$ values. 

Also, we note that the algorithm refines the values of $\bm{\eps}(i)$ throughout the iterations.
As the algorithm progresses, $\bm{\eps}(i)$ is iteratively updated towards the real and observed $1-\bm{{\rho}}(i)$, computed using the current values $\bm{\bar{f}}(i), \bm{\bar{\mu}}(i)$. This has the effect of making the upper bound $\E[\bar{D}^{k,r}_{uv}(\bm{f}, \bm{\mu}, \bm{\eps})]$ converge to its real value $\E[D^{k,r}_{uv}(\bm{f}, \bm{\mu}, \bm{\eps})]$, whose formulation is non-convex.





%
%
%
%
%
%

We have presented an algorithm that decomposes and solves the optimization problem iteratively in the continuous domain. However, to obtain an admissible solution for $\mathcal{P}$, the flow variables are required to be integer. This necessitates the use of a rounding strategy. In this work, we adopt the same rounding technique used in IDAGO, a bi-criteria optimization algorithm originally introduced in \cite{idago}. IDAGO was developed to address multi-commodity flow placement and resource allocation, followed by integer rounding of flow variables on the same cloud-augmented graph. The IDAGO rounding technique is efficient, requiring the heavy convex optimization step to be executed only once.
The IDAGO rounding procedure generates multiple candidate embeddings, where each embedding consists of a set of rounded flow variables that satisfy flow conservation. Probabilities are assigned to each candidate embedding, and the final embedding is selected through randomized sampling based on these probabilities. 
IDAGO comes with formal guarantees on the quality of the solution, expressed in terms of its approximation ratio.

Eventually, SPARQ finds a solution $\mathcal{S}$ with integer ${f}^k_{uv}$ variables which is returned as the final solution of problem $\mathcal{P}$. In the following section, we apply the whole modeling, decomposition and algorithm pipeline discussed in the previous sections to our experimental setup and simulations.




\section{Results}
In this section, we present the experimental setup and the results we obtained. 
In all experiments, we assessed our queue modeling approach against the private delay model in \cite{idago}. To strengthen the allocation given by the private delay model, rates $\mu^r_{uv}, \mu^{k,r}_{uv}$ are deliberately over-allocated by a constant $\alpha \geq 1$.
We will observe that, no matter the multiplication constant $\alpha$, the system with the private delay model produces either non-valid or sub-optimal solutions: either because the allocated resources are too low to satisfy the latency constraints or because over-allocation generates a sub-optimal and unnecessarily costly solution.

In all experiments, we apply SPARQ to find a solution for the placement, routing and resource allocation. We then use the values of the variables and compute values such as the latency and the objective cost.

\subsection{Experiment A}
\begin{figure}[t]
    \centering
    \begin{subfigure}[b]{0.98\linewidth}
        \centering
        \includegraphics[width=0.8\linewidth]{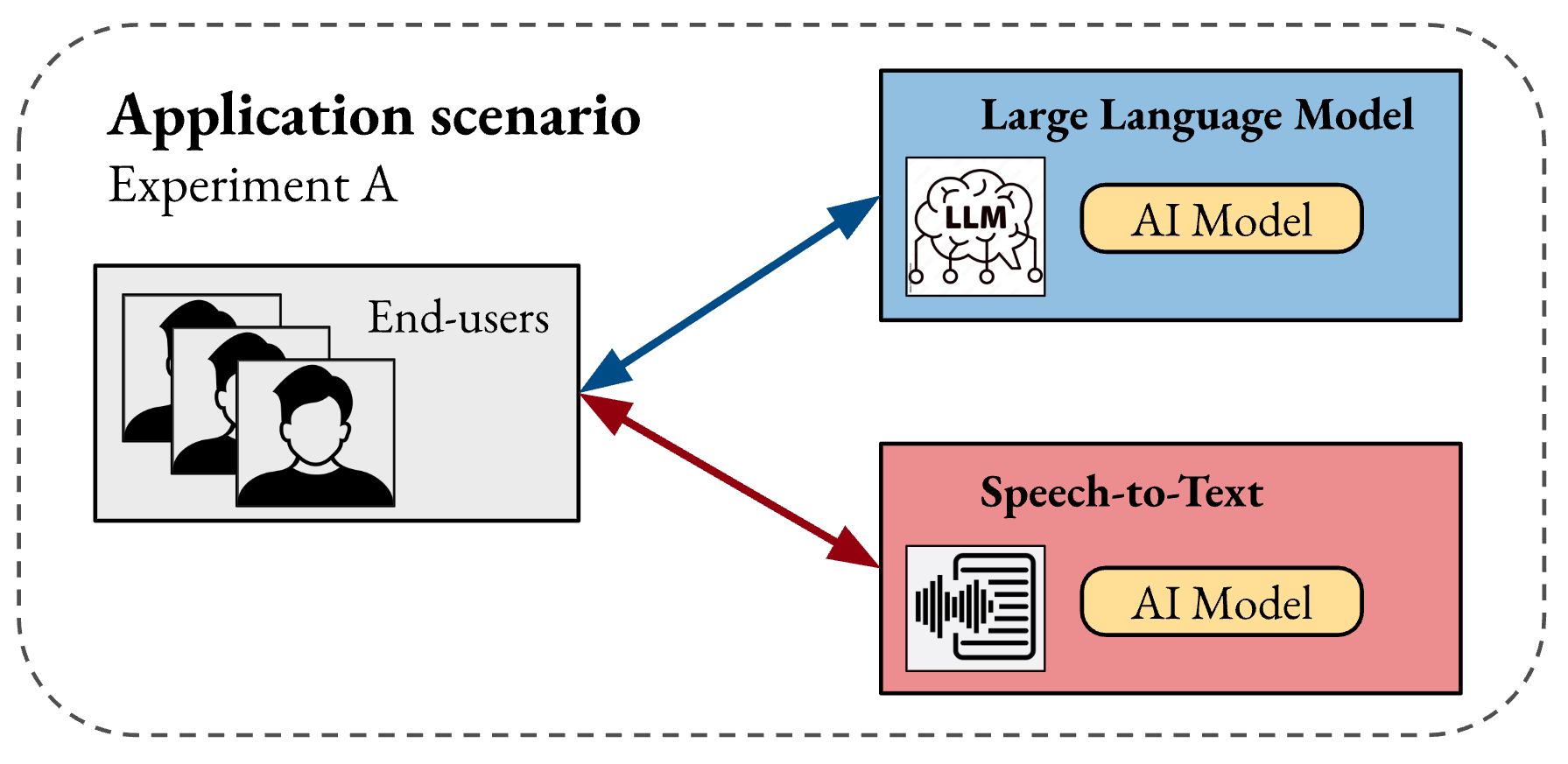}
        \caption{Application overview of experiment A. A pool of users accesses different AI models to achieve different tasks.}
        \label{fig:exp_a_app_graph}
        \vspace{0.3cm}
    \end{subfigure}
    \begin{subfigure}[b]{0.98\linewidth}
        \centering
        \includegraphics[width=0.8\linewidth]{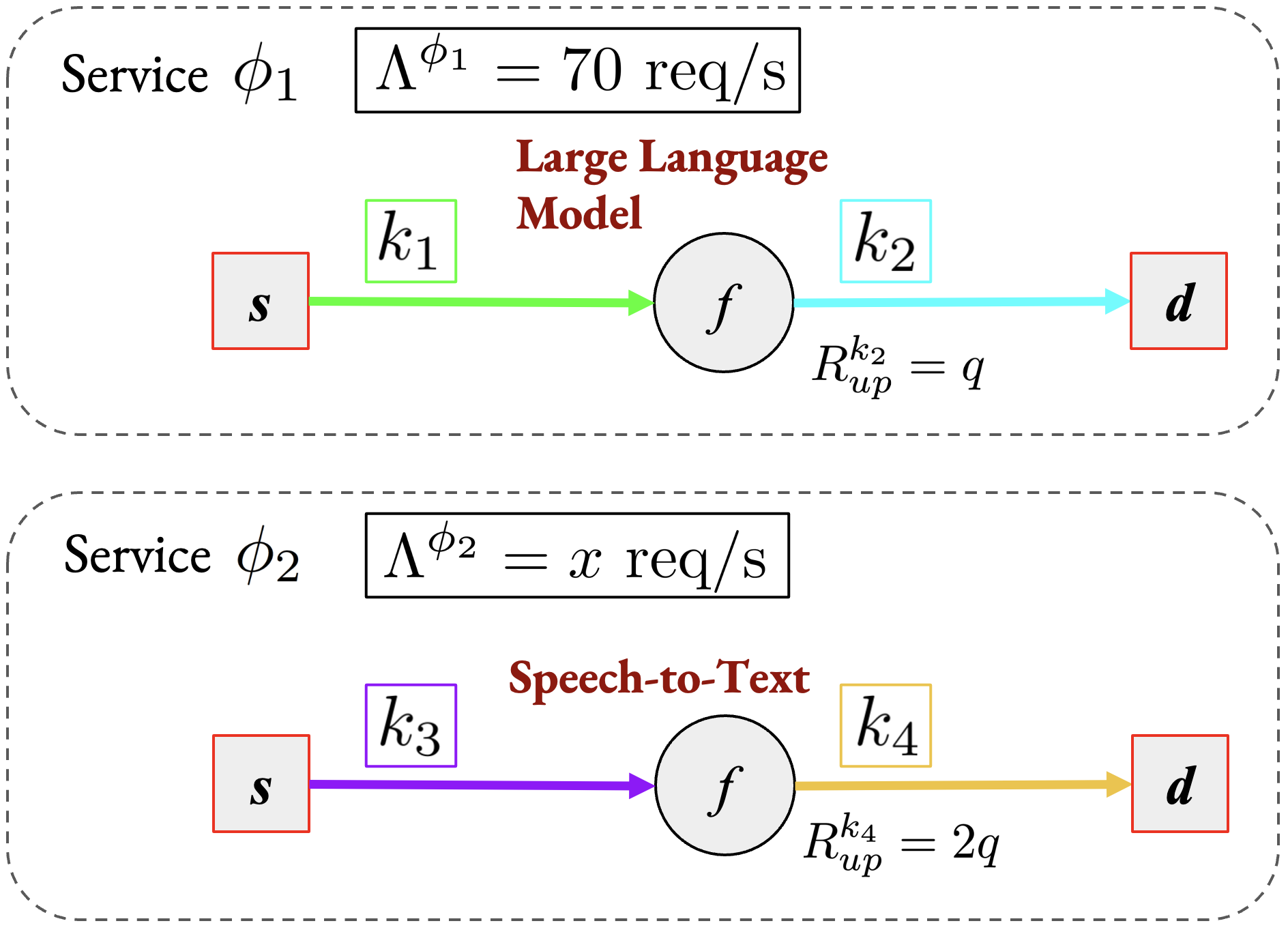}
        \caption{Service graph of experiment A, consisting of two services with only two commodities each, i.e. one input and one output commodity with a single function.}
        \label{fig:exp_a_service_graph}
        \vspace{0.3cm}
    \end{subfigure}
    \begin{subfigure}[b]{0.98\linewidth}
        \centering
        \fbox{\includegraphics[width=0.8\linewidth]{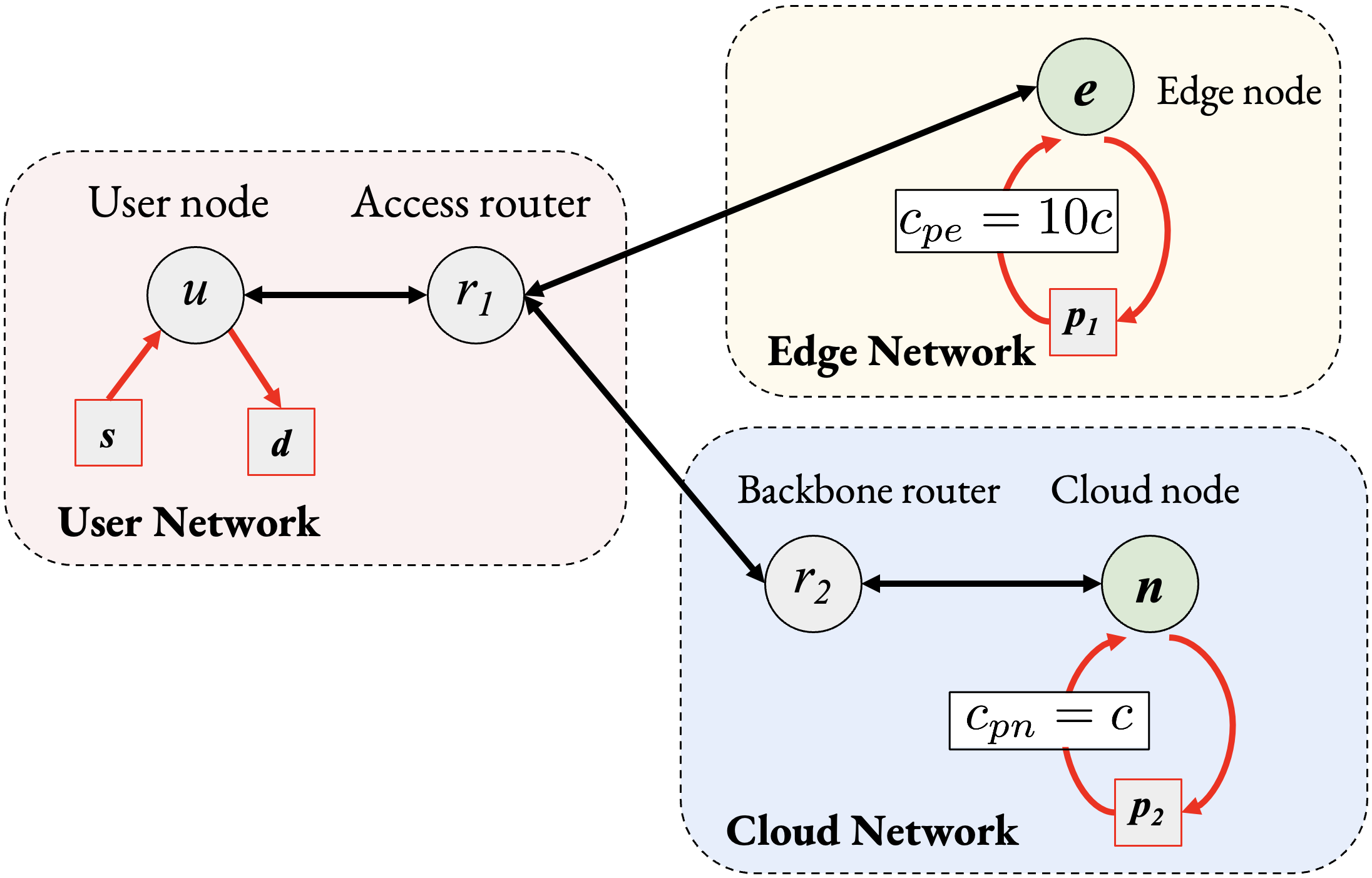}}
        \caption{Augmented graph of experiment A, consisting of a network with two computation nodes, having different computational costs.}
        \label{fig:exp_a_network_graph}
    \end{subfigure}
    \caption{Experiment A application, augmented graph and service graph.}
    \label{fig:exp_a_combined}
\end{figure}

The first experiment involves a small network with a pool of users and two available AI services, reported in Fig. \ref{fig:exp_a_combined}. The goal of this experiment is to determine the optimal computation placement, flow routing and resource allocation across the edge-cloud network. An overview of the application is reported in Fig. \ref{fig:exp_a_app_graph}. We modeled this experiment starting from the service graph, reported in Fig. \ref{fig:exp_a_service_graph}. We consider two AI services acting on the network: a Speech-to-text model and a Large Language Model, respectively indicated as $\phi_1, \phi_2$, In our experiments we modeled $\phi_2$ to require double the amount of processing than $\phi_1$, namely $q$. Given the AI and heavyweight nature of the services, both of them are modeled following the \SR model. Both services have network node $u$ as respective source and destination.
In Fig. \ref{fig:exp_a_network_graph} is reported the augmented graph of the edge-cloud network. It consists of a user node $u$, a cloud computational node $n$ and an edge computational node $e$, connected by routers. We modeled the cloud processing cost $c_{pn}$ as a parameter $c$, and we set the edge processing cost to be $c_{pe}=10c$. The cost coefficients are generic and tunable, allowing the model to reflect operator or application preferences. For instance, lowering edge costs favors edge processing, while higher values naturally bias the computation toward the cloud.



\begin{figure}[t]
    \centering
    \includegraphics[width=0.9\linewidth]{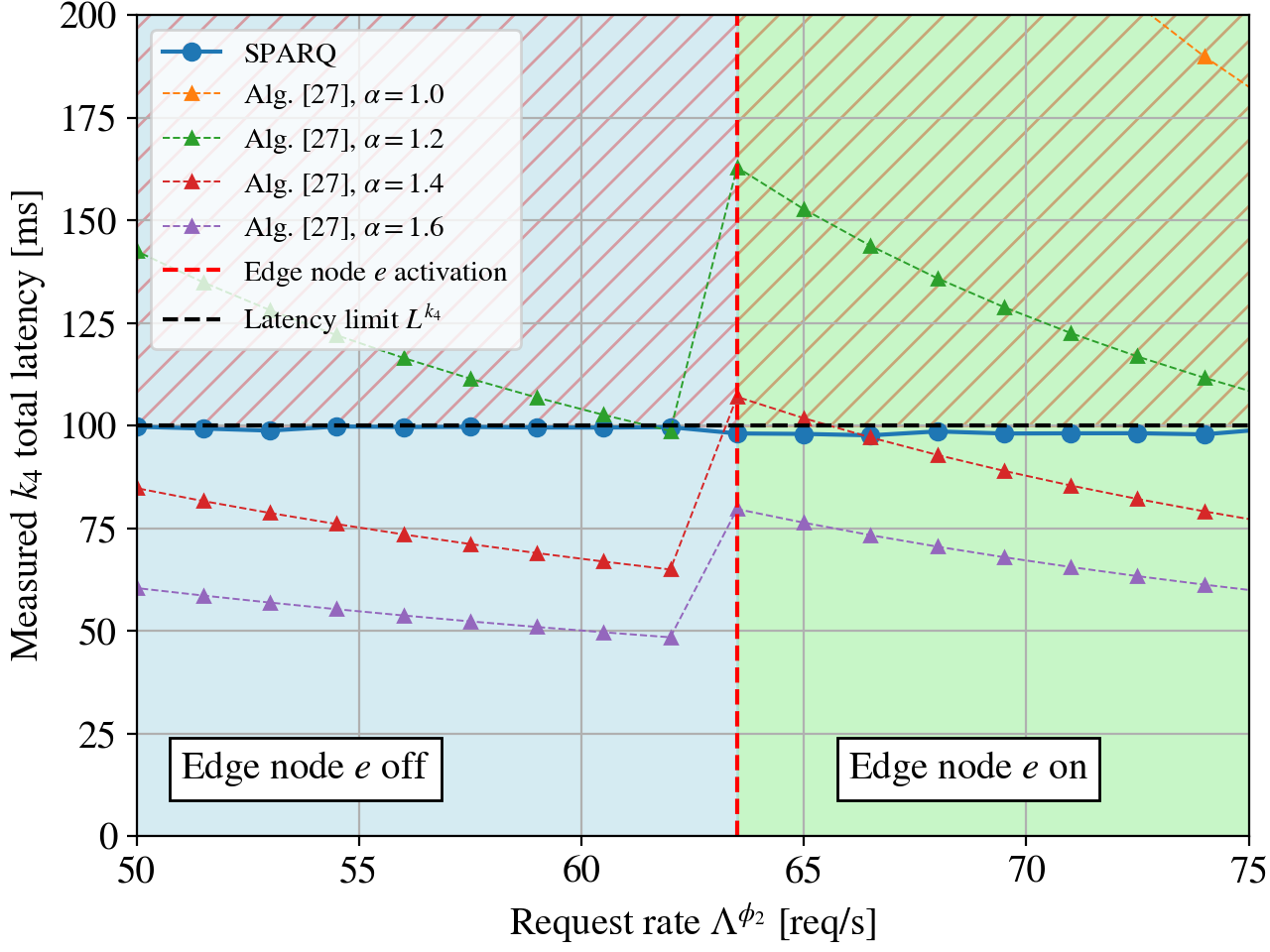}
    \caption{Measured latency of output commodity $k_4$ on rate $\Lambda^{\phi_2}$ change for experiment A. The dashed black horizontal line is the maximum latency limit $L^{k_4}$ and the vertical red line is the activation of processing of node $e$. The upper area of the graph, shaded with red diagonal lines, violates the latency constraints.}
    \label{fig:exp_a_latency_zoom}
\end{figure}

In this experiment, we fixed the request arrival rate of the first service $\Lambda^{\phi_1}$ and we increased the rate of the second service $\Lambda^{\phi_2}$. The two output commodities $k_2, k_4$ are constrained to a maximum latency of $L^{k_2}=L^{k_4}=100$ms.
In Figure \ref{fig:exp_a_latency_zoom} is reported the measured latency of $k_4$.
When the value of $\Lambda^{\phi_2}$ is low, the system performs all computations on node $n$, as it is cheaper to operate. However, as the rate $\Lambda^{\phi_2}$ increases beyond a certain threshold, node $n$ can no longer handle the processing of both commodities, and the system is forced to activate node $e$. The rate at which this transition occurs is indicated by the vertical red line in the figure. 
The reported latency is the real, measured, a-posteriori latency, which is \textit{not} directly optimized by the algorithm since it is non-convex. When measured after the optimization algorithm has set the variables values, it remains very close to the boundary, hence to the optimality.
We observe that the private delay model solution struggles to stay below the $L^{k_4}$ limit, caused by the non-balanced distribution of requested service $R^{k,r}_{uv}$ on the same shared resource. For the latency to exceed $L^{k_4}$, the rates must be multiplied by a constant $\alpha$ greater than $1.4$ approximately. However, such solutions are sub-optimal, being well below the latency boundary, resulting in a cost-inefficient allocation, as shown in the following.

\begin{figure}[t]
    \centering
    \includegraphics[width=0.9\linewidth]{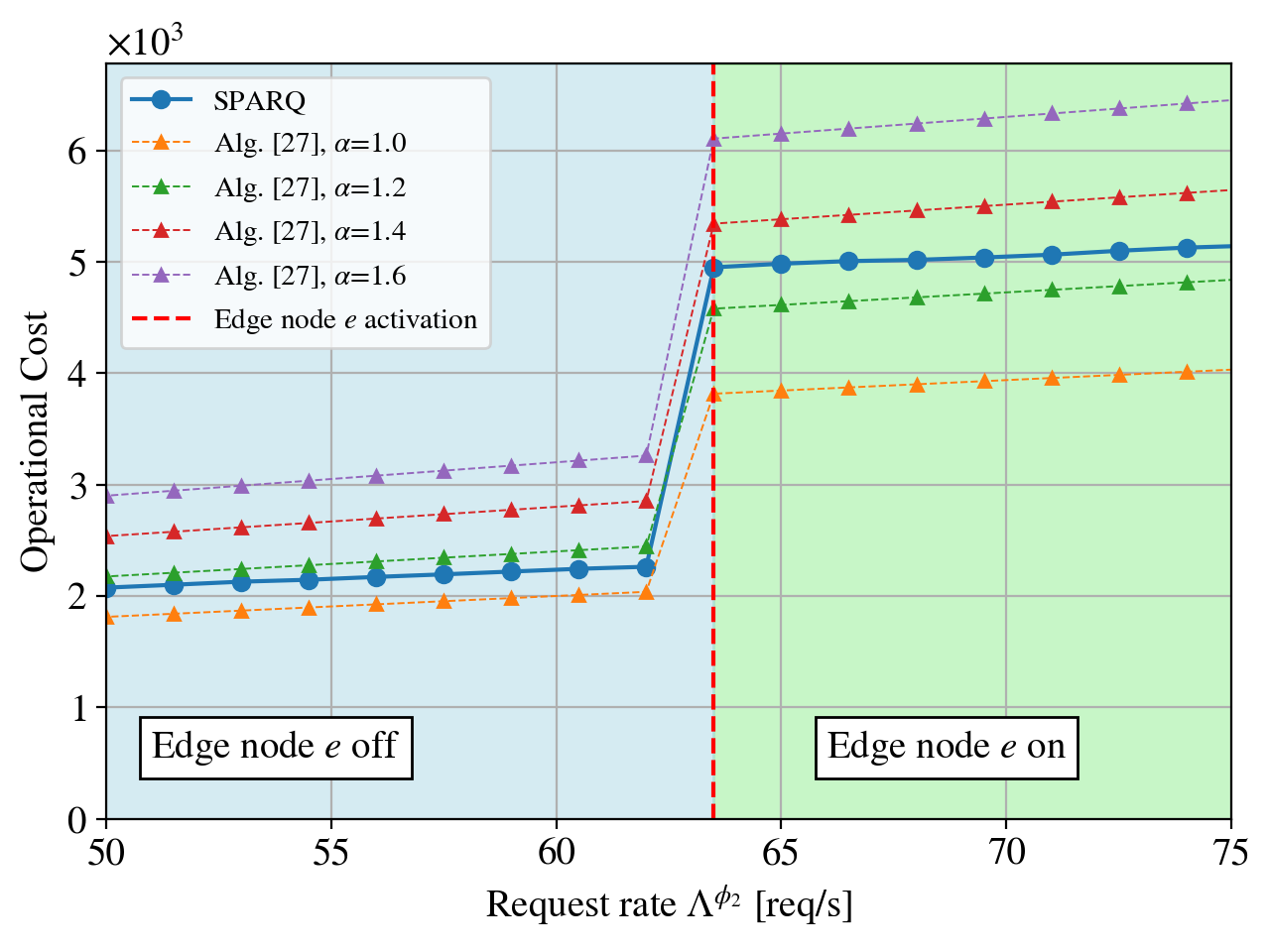}
    \caption{Operational cost of experiment A on change of request rate $\Lambda^{\phi_2}$.}
    \label{fig:exp_a_obj}
\end{figure}

\begin{figure}[t]
    \centering
    \includegraphics[width=0.98\linewidth]{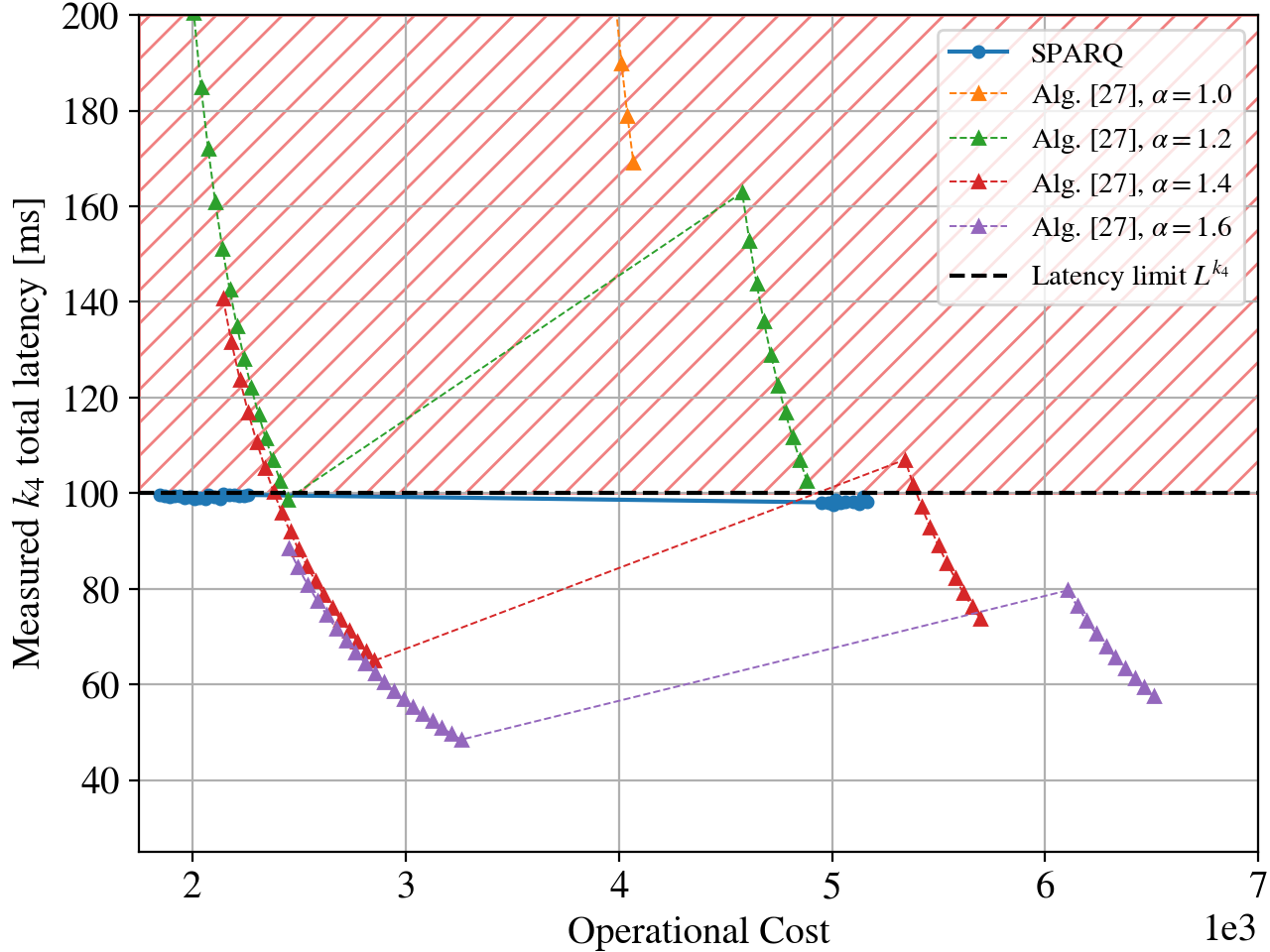}
    \caption{Trade-off curves of Experiment A for the different solutions. With equal latency, our solution has the lowest operational cost which stays below the maximum latency limit $L^{k_4}$.}
    \label{fig:exp_a_tradeoff}
\end{figure}

Fig. \ref{fig:exp_a_obj} shows the behavior of the objective function. We remark that the objective function represents the operational cost of the placement, routing and resource allocation of the deployment. 
In the private delay model, the cost increases as the allocation multiplier $\alpha$ increases. Private delay model solutions with $\alpha \in \{1.4, 1.6\}$ achieve a higher cost compared with SPARQ. Solutions with $\alpha\in\{1.0, 1.2\}$ yield a lower cost than SPARQ but produce latencies exceeding the threshold $L^{k_4}$ by a significant margin.
The cost-efficiency of the solution can be observed in Fig. \ref{fig:exp_a_tradeoff}, presenting the trade-off curves of the different solutions on the cost-latency space. In this experiment, the constant latency limit bisects the plane into two areas which contain solutions with measured latency below and above 100ms respectively, the latter being unfeasible for the problem. Again, we observe that our solution lies close to the feasible boundary, indicating the scarce resource over-allocation carried out by our solution. From the trade-off curves we can observe that with equal latency we obtain the lowest cost compared to the other solutions.

\subsection{Experiment B}
\begin{figure}[t]
    \centering
    \begin{subfigure}[b]{\linewidth}
        \centering
        \includegraphics[width=\linewidth]{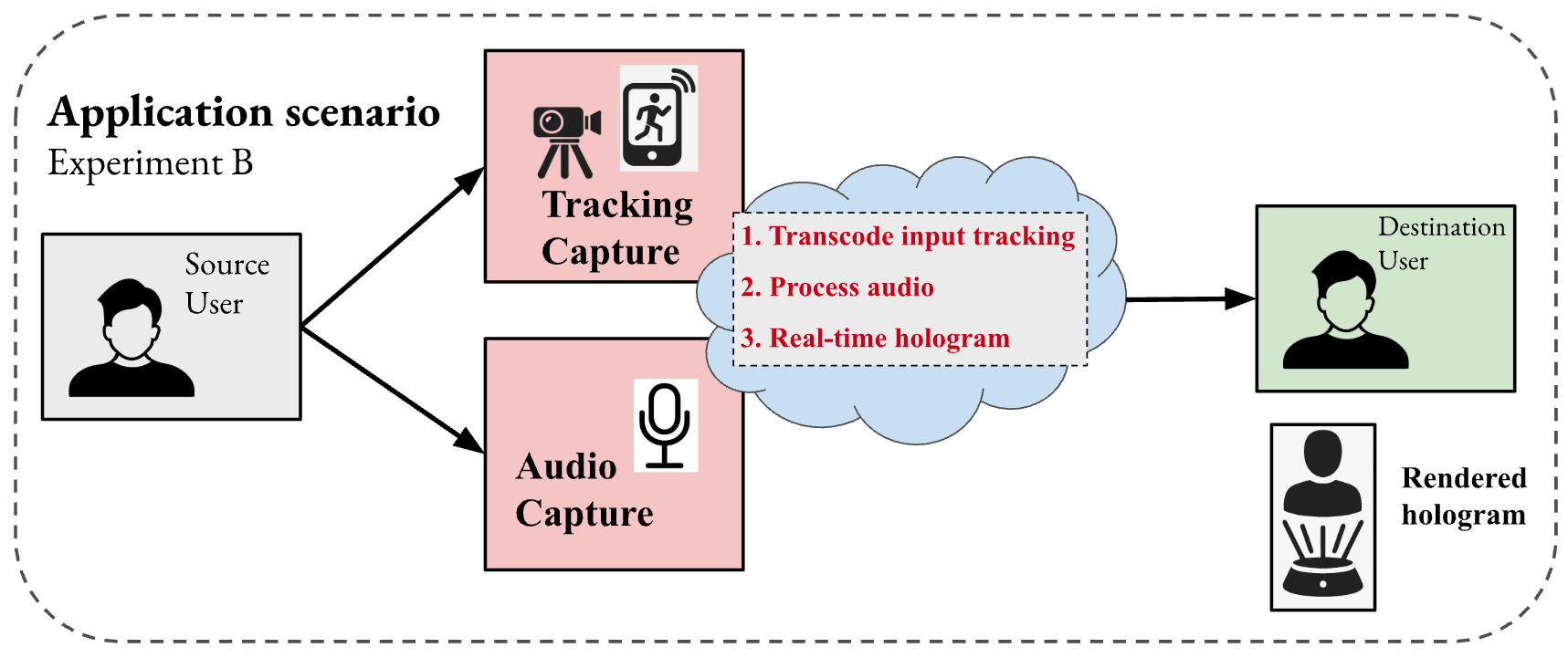}
        \caption{Application overview of FANTASIA \cite{fantasia_app}.}
        \label{fig:exp_b_setup}
        \vspace{0.3cm}
    \end{subfigure}
    \begin{subfigure}[b]{\linewidth}
        \centering
        \includegraphics[width=\linewidth]{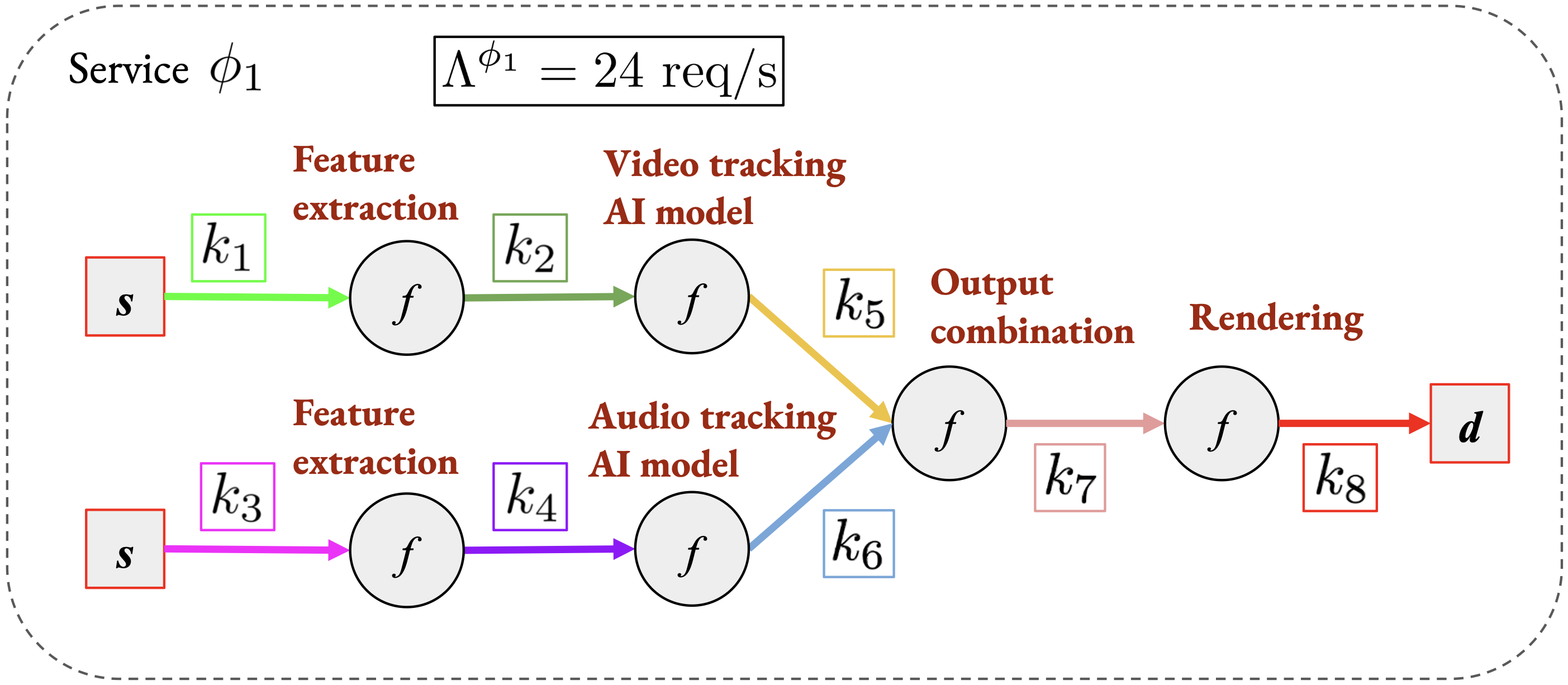}
        \caption{Application service graph of FANTASIA. The commodity $k_8$ represents the final rendered data stream.}
        \label{fig:exp_b_service_graph}
        \vspace{0.3cm}
    \end{subfigure}
    \begin{subfigure}[b]{\linewidth}
        \centering
        \includegraphics[width=\linewidth]{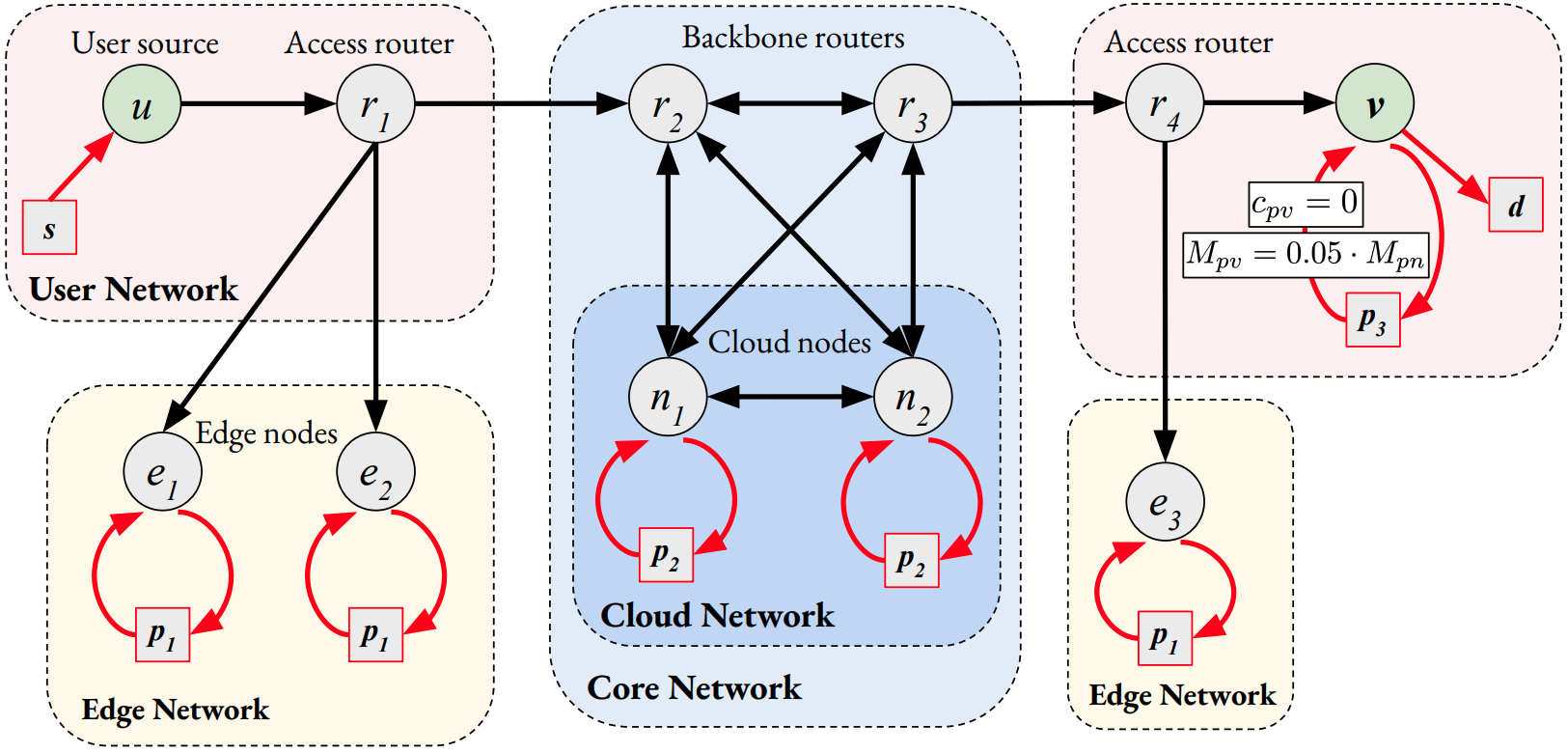}
        \caption{Augmented graph of experiment B. The user equipment, represented by node $v$, has a computation node modeled with 0 associated cost and a capacity which is $5\%$ of the capacity of the other nodes in the network.}
        \label{fig:exp_b_network_graph}
    \end{subfigure}
    \caption{Experiment B application, service graph and network graph.}
    \label{fig:exp_b_combined}
\end{figure}

In this experiment, we investigate an application involving an Augmented Reality (AR) communication system. This application is based on the FANTASIA framework \cite{fantasia}, which has already been implemented in other applications \cite{fantasia_app}. The modeling is shown in Fig. \ref{fig:exp_b_combined}. Specifically, the overview of the application is shown in Fig. \ref{fig:exp_b_setup}. The system is designed as a provider of AR communication, where the user captures video, motion tracking, and audio. The application is responsible for managing raw data streams, including audio, video, and tracking information. Afterward, the system performs feature extraction on the captured data, passes it through AI models for various tasks such as recognition, reconstruction, and other processing purposes. The outputs from these models are then combined and rendered onto a final AR/VR device for the destination user, creating a real-time holographic experience. This is summarized in the application service graph in Fig. \ref{fig:exp_b_service_graph}. The commodities requiring process by AI models, are modeled following the $\SR$ model, while the others follow the $\GR$ model. The network graph in Fig. \ref{fig:exp_b_network_graph} shows the augmented graph of the system, consisting of multiple nodes and paths connecting the sources of the source user's streams to the destination user's node $v$.
We observe that many nodes in the network possess computational capabilities, including those in edge networks close to users as well as cloud nodes. In particular, the destination node $v$ also has processing capabilities that can be utilized to process commodities.

In this experiment, we analyze the optimal service allocation from the application provider's perspective, focusing on minimizing the economic cost. Here, the provider is faced with the decision of how to allocate computational tasks, having two options:
\begin{enumerate}
    \item processing the video streams on edge and/or cloud resources incurring in operational costs;
    \item offloading computations to the user's end-device, incurring in no operational costs, but considering that the user’s equipment has limited computational capabilities.
\end{enumerate}
In particular, the provider may decide whether to offload the final rendering operation, represented by commodity $k_8$, to the user's equipment, provided that latency and capacity constraints are met. We modeled this environment as follows. The process of producing the output commodity $k_8$ can be offloaded to $v$ with an associated cost $c^r_{pv}=0$ and $v$'s processing capacity is constrained to 5\% of the capacity of other computing nodes.

\begin{figure}[t]
    \centering
    \includegraphics[width=0.98\linewidth]{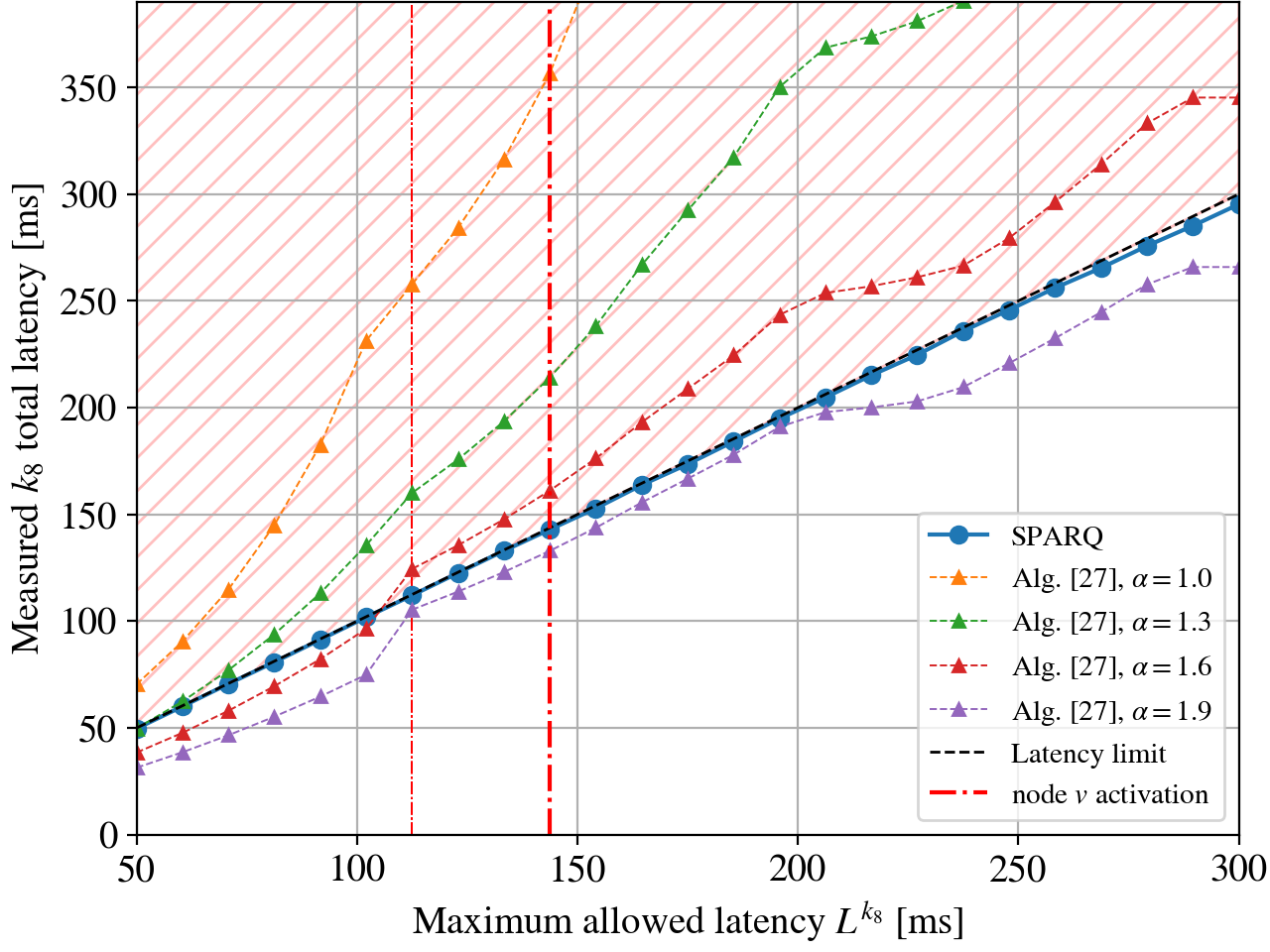}
    \caption{Measured latency experienced by the application, subject to latency constraints.}
    \label{fig:exp_b_lat_real}
\end{figure}

In this experiment, we varied the maximum allowed latency $L^{k_8}$ and observed the system's response. Fig. \ref{fig:exp_b_lat_real} shows the total latency experienced by the application.
As in the previous experiment, the considered latency is the real, measured latency which is not directly optimized by the system. The black dashed line in Fig. \ref{fig:exp_b_lat_real} indicates the latency limit. An optimal solution should align with this boundary, and our solution consistently remains very close without exceeding it. In our solution, when $L^{k_8}$ exceeds approximately 140ms, the system activates processing on user node $v$, indicated by the rightmost vertical red line in the figure. 
Below this threshold, the system recognizes that even fully utilizing the computational resources of node $v$, it would not meet the latency constraint. Consequently, node $v$ is disabled, and the computation is shifted to other nodes. 
In contrast, the private delay model solutions are less conservative and continue to use node $v$ for lower and more stringent values of $L^{k_8}$, indicated by the leftmost vertical red line in the figure, which results in exceeding the latency constraint. 


\begin{figure}[t]
    \centering
    \includegraphics[width=0.8\linewidth]{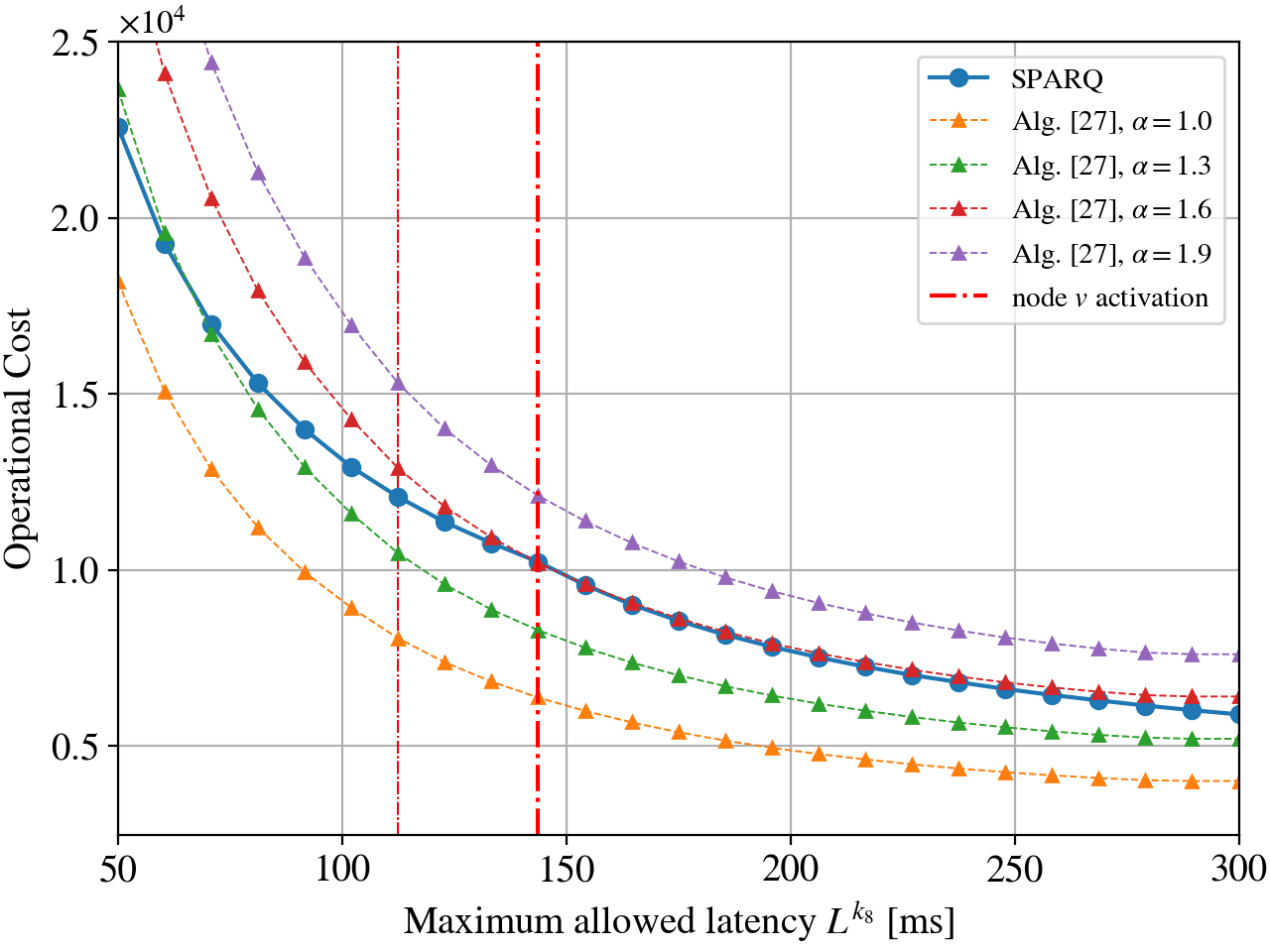}
    \caption{Operational cost of the solution on the variation of the maximum allowed latency $L^{k_5}$.}
    \label{fig:exp_b_cost}
\end{figure}

Fig. \ref{fig:exp_b_cost} shows the cost associated to the solutions.
Combining the results from figures \ref{fig:exp_b_lat_real}, \ref{fig:exp_b_cost}, it can be seen that solutions with $\alpha\in\{1.0, 1.3\}$ always exceed the threshold, violating the latency constraint. The solution with $\alpha=1.6$ starts below the latency limit and costing more than our solution, and then violates the latency limit when it activates node $v$ for the computation. Finally, the solution with $\alpha=1.9$ stays always below the latency requirement but it allocates more resources respect to our solution, leading to sub-optimal solutions with unnecessarily low latency and higher costs.

\begin{figure}
    \centering
    \includegraphics[width=0.9\linewidth]{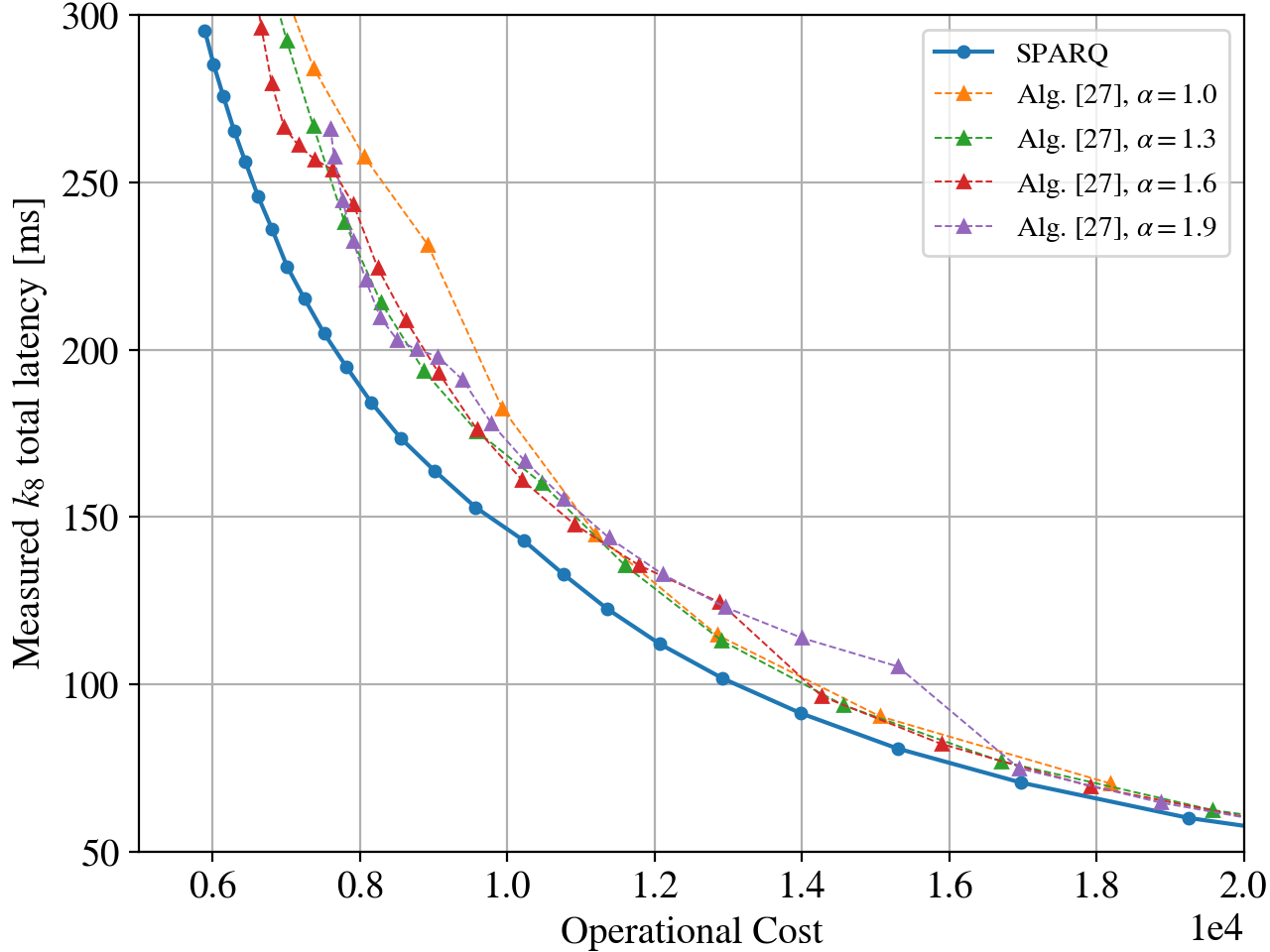}
    \caption{Trade-off curves of Experiment B in the cost-latency space for the different solutions. Our solution attains the most efficient placement and resource allocation compared to the others.}
    \label{fig:exp_b_tradeoff}
\end{figure}

Fig. \ref{fig:exp_b_tradeoff} illustrates the trade-off in the cost-latency space for the different solutions. In this experiment, the maximum latency constraint is variable, meaning that it does not strictly divide the plane into feasible and infeasible regions as Experiment A. Nevertheless, the curves demonstrate that our proposed approach consistently achieves the lowest cost for a given latency, even though some solutions may fall outside the feasible range. Furthermore, the comparison curves are closely aligned, indicating that the efficiency of these alternative solutions is similar. In contrast, our solution exhibits a clear separation from the others, highlighting its higher efficiency by a notable margin.

\section{Conclusions}
In this work, we tackle the optimization of placement, routing, and communication-computation resource allocation of AI-intensive applications over edge-cloud architectures. 
We propose a novel modeling approach that by combining two key resource models, Guaranteed-Resource (GR) and Shared-Resource (SR) models, provides a more accurate representation of how modern applications are executed on physical hardware, particularly in the context of resource-intensive AI workloads. The \GR model ensures dedicated resources for each task, while the \SR model accounts for the shared nature of certain resources, where delays are influenced by different tasks running concurrently.

To model the resulting intricate resource-delay dynamics, we introduce a queue-based methodology that uses a mixture of M/G/1 and M/M/1 queues to accurately capture the non-linear relationship between resource allocation and service delays.


Given the non-convexity of the resulting optimization problem, we develop a set of approximation and convexification techniques that decompose the 
problem into two tractable sub-problems. The proposed iterative algorithm, termed SPARQ, jointly optimizes service placement, routing, and resource allocation under non-linear delays. Experimental results demonstrate the effectiveness of our approach over traditional methods. In particular, SPARQ produces a solution of lower cost, while maintaining latencies under the maximum threshold, even in the presence of unbalanced service requests.

To conclude, we remark that the generality of the resource-delay model presented in this paper makes it suitable for many emerging applications and computing environments. 
A direction for future work is to apply this model in production environments, such as Amazon AWS, for the orchestration of real agentic-AI applications. 



\bibliographystyle{ieeetr}
\bibliography{bibliography}

\end{document}